\UseRawInputEncoding
\RequirePackage{fix-cm}

\documentclass[smallextended]{svjour3}       % onecolumn (second format)
\smartqed 

\usepackage{graphicx}
\usepackage{amssymb}
\usepackage{hyperref}
\usepackage{xcolor}
\usepackage{float}
\usepackage{natbib}
\usepackage{textcomp}
\usepackage{gensymb}
\usepackage{booktabs}       % professional-quality tables
\usepackage{siunitx}    % For number formatting
\usepackage{pgfplots}   % For advanced TikZ/PGF matrix and plot features
\usepackage{multirow}   % For table multirows
\usepackage{threeparttable} % For table notes
\usepackage{bm}
\usepackage{mathtools}
\usepackage{physics}
\usepackage{tikz}
\usetikzlibrary{arrows.meta,positioning,calc,fit,shapes.geometric,shapes.misc,matrix}
\usepackage{adjustbox} % for \resizebox
\usepackage{graphicx}
\DeclareGraphicsExtensions{.pdf,.png,.jpg}
\usepackage{booktabs}
\usepackage{siunitx}
\usepackage{amsmath}

\allowdisplaybreaks[2]  % 0..4 (higher = more willing to break)

\begin{document}

\title{The Shape of Markets: Machine learning modeling and Prediction Using 2-Manifold Geometries}


\author{Panagiotis G. Papaioannou \and Athanassios N. Yannacopoulos}


\institute{
           Panagiotis G. Papaioannou \at
           Department of Statistics, Athens University of Economics \& Business, Athens, GR 10434; Stochastic Modelling and Applications Laboratory, Athens University of Economics \& Business, Athens, GR \\
               \email{ppapaioannou@aueb.gr}    
              \and
              Athanassios N. Yannacopoulos \at
              Department of Statistics, Athens University of Economics \& Business, Athens, GR 10434; Stochastic Modelling and Applications Laboratory, Athens University of Economics \& Business, Athens, GR \\
              \email{ayannaco@aueb.gr}  
}

\date{Received: date / Accepted: date}

\maketitle

\begin{abstract}
We introduce a \textit{Geometry-Informed Model} for financial forecasting by embedding high-dimensional market data onto constant-curvature 2-manifolds. Guided by the uniformization theorem,\cite{Thurston1997}, we model market dynamics as Brownian motion on spherical ($S^2$), Euclidean ($R^2$), and hyperbolic ($H^2$) geometries. We further include the torus ($T²$), a compact, flat manifold admissible as a quotient space of the Euclidean plane-—anticipating its relevance for capturing cyclical dynamics,\cite{doCarmoCurves}. Manifold learning techniques infer the latent curvature from financial data, revealing the torus as the best-performing geometry. We interpret this result through a macroeconomic lens: the torus's circular dimensions align with endogenous cycles in output, interest rates, and inflation described by IS-LM theory \cite{Hicks1937}. Our findings demonstrate the value of integrating differential geometry with data-driven inference for financial modeling,\cite{LopezDePrado2018AFML,CapponiLehalle2023}.
% keywords can be removed

\end{abstract}

\section{Introduction}
Financial markets are characterized by their complex, non-linear behaviors, posing significant challenges for modeling and prediction. While traditional approaches to market analysis rely heavily on statistical tools and Euclidean assumptions, these methods often fail to account for deeper geometric structures inherent in the data. A novel perspective can be gained by incorporating concepts from differential geometry, particularly the classification of 2-manifolds into three geometries: spherical, Euclidean, and hyperbolic.\cite{Thurston1997}

The geometrization theorem, which underpins this classification, provides a foundational framework for understanding how spaces can be decomposed into simpler geometric components,\cite{Perelman2002}. This insight, central to modern differential geometry, suggests that many complex systems, including financial markets, may exhibit characteristics aligning with one or more of these geometric types. As in physics and network science, the widespread application of geometric methods and their potential in financial market analysis is actively explored under active research,\cite{CapponiLehalle2023,SimonianFabozzi2019,LopezDePrado2018AFML,LopezDePrado2020MLAM,NoguerAlonso2021FinEAS}. Recent studies explore the application of manifold learning techniques to financial market analysis and time series forecasting. These methods aim to extract low-dimensional representations of complex, high-dimensional data, revealing intrinsic structures and patterns (\cite{Y2017Nonlinear}; \cite{G2018Nonlinear}; \cite{Y2020Manifold}). In \cite{Jansen2023} several trading practice applications and concepts are outlined, as well. Manifold learning approaches have been used for early warning systems in financial markets (\cite{Y2017Nonlinear}; \cite{G2018Nonlinear}), phase space reconstruction of financial time series (\cite{Y2014Manifold}), and visualization of market states (\cite{Y2020Manifold}). Researchers have also developed novel algorithms, such as information metric-based manifold learning (\cite{Y2017Nonlinear}) and kernel entropy manifold learning (\cite{Y2014kernel}), to improve the accuracy of financial analysis and prediction. Beyond finance, manifold learning techniques have been applied to various time series analysis tasks, including electroencephalography signal analysis (\cite{P2018Multivariate}) and forecasting of high-dimensional time series (\cite{P2021Time}), demonstrating their versatility in capturing complex dynamics across different domains.\cite{Perelman2002}

Brownian motion, a stochastic process traditionally studied in Euclidean spaces, is a powerful tool for modeling random dynamics. Extending Brownian motion to spherical and hyperbolic geometries (\cite{Hsu2002,IkedaWatanabe1989}) enables the representation of processes influenced by curvature and global topological features. Studies have explored diffusion on spheres (\cite{Gomez2021geometrical}; \cite{M2000Brownian}) and hyperbolic spaces (\cite{L2007Hyperbolic}, \cite{Hsu2002StochasticAO}), as well as fractional Brownian motion in both geometries (\cite{J2005Spherical}). The effects of curvature on Brownian motion have been investigated using Riemann normal coordinates (Castro-Villarreal, 2010) and the Smoluchowski equation (\cite{Pavel2014Intrinsic}). Research has shown that positive curvature slows diffusion, while negative curvature accelerates it (\cite{Pavel2010Brownian}). The concept of Brownian motion has been generalized to metric spaces of constant curvature (\cite{S1981Brownian}). Additionally, studies have examined the hydrodynamics of curved membranes and their impact on particulate mobility (\cite{Henle2010Hydrodynamics}), providing insights into diffusion on biological structures like vesicles and membrane tethers.\cite{Hsu2002,IkedaWatanabe1989}

Financial markets exhibit complex, nonlinear dynamics that may be better understood through non-Euclidean geometric frameworks (\cite{E2006Geometry}; \cite{Ilinski1999Gauge};\cite{Ilinski2001,Emami2021,KellerRessel2021}). These markets can be modeled using projective geometry, fiber bundles, and fractal geometry to capture their intricate structures and behaviors (\cite{E2006Geometry}; \cite{Ilinski1999Gauge}; \cite{B2004MIS}; \cite{Lipton2001FX,Lipton2018SelectedWorks}). Researchers have developed methods to incorporate non-Euclidean geometric information into filtering and machine learning algorithms, improving their accuracy in financial applications (\cite{Anastasis2017Geometric}). Network geometry measures, such as discrete Ricci curvatures, can be used to analyze market instability and systemic risk (\cite{Samal2021Network}). The concept of "dark volatility," a hidden factor influencing market behavior, has been explored using Einstein warped product manifolds (\cite{Pincak2023possible}). These geometric approaches provide powerful tools for modeling asset behavior, forecasting volatility, and evaluating investment risks with greater precision (\cite{B2004MIS}; \cite{Humphrey2011Financial}).(\cite{Ilinski2001,KellerRessel2021,Emami2021})

This paper introduces a new methodology that leverages the interplay between geometry, stochastic processes, and machine learning to analyze and predict financial market behavior. The proposed approach consists of three steps:

\begin{enumerate}
   \item \textbf{Brownian Motion Construction}: Simulate Brownian motions within spherical, Euclidean, and hyperbolic geometries to represent market dynamics.
    \item \textbf{Geometry Inference}: Apply local gaussian curvature analysis, to infer the underlying geometry from observed trajectories.
    \item \textbf{Market Prediction}: Use the inferred geometry to inform predictions of future market behavior, leveraging the geometric context to enhance accuracy.
\end{enumerate}

The key idea is to treat financial market dynamics as trajectories embedded in geometrically diverse spaces. This approach allows us to move beyond traditional Euclidean-based analyses, uncovering the "shape" of market data and providing insights into its structural properties. For example, spherical geometry might correspond to cyclic or periodic market behaviors, Euclidean geometry to stable trends, and hyperbolic geometry to highly volatile and chaotic dynamics. By connecting geometric theory with practical market analysis, this work offers contributions in both theory and application. Theoretically, it bridges the abstract mathematical classification of 2-manifolds with stochastic processes and financial modeling. Practically, it provides a new tool for market prediction that is particularly well-suited for analyzing non-linear and non-stationary systems.

This study aims to demonstrate that the geometric classification of trajectories, combined with stochastic modeling and machine learning, can significantly improve the understanding and prediction of financial markets. The results highlight the power of geometry-informed models in capturing complex behaviors, offering new avenues for research and application in finance and beyond.

\section{Concept and Scope: From Linear Orthogonality to Geometric Structure}
\label{sec:concept_scope}

This study bridges the gap between linear dimensionality reduction and non-linear geometric modeling. Our approach rests on three conceptual pillars: the critical distinction between orthogonality and independence in PCA, the operational necessity of identifying a low-dimensional manifold via the Uniformization Theorem, and the rigorous construction of tradable eigenportfolios.

\subsection{Dimensionality Reduction and Tradable Eigenportfolios}
We employ an expanding window Principal Component Analysis (PCA) to extract the latent drivers of the market structure. Let $\mathbf{R}_t \in \mathbb{R}^{N}$ denote the vector of standardized log-returns for $N$ assets at time $t$. The empirical covariance matrix $\Sigma_t$ is decomposed via Spectral Decomposition:
\begin{equation}
    \Sigma_t = \mathbf{V}_t \mathbf{\Lambda}_t \mathbf{V}_t^\top
\end{equation}
where $\mathbf{V}_t$ is the matrix of eigenvectors (loadings) and $\mathbf{\Lambda}_t$ the diagonal matrix of eigenvalues.

To ensure the resulting portfolios are \textit{tradable} and free from look-ahead bias, we construct the weights for the eigenportfolios at time $t$ as derived from the eigenvectors computed at time $t-1$. Thus, the return $F_{k,t}$ of the $k$-th eigenportfolio is realized as:
\begin{equation}
    F_{k,t} = \mathbf{R}_t \cdot \mathbf{v}_{k,t-1}
\end{equation}
This lag ensures that the portfolio composition is known at the close of $t-1$, making the strategy executable in a realistic trading environment \cite{Avellaneda2010, Jolliffe2002}.

We retain the first three principal components ($k=1,2,3$), constructing a 3-dimensional embedding $\mathbf{X}_t = (F_{1,t}, F_{2,t}, F_{3,t})^\top \in \mathbb{R}^3$. Descriptive analysis of our dataset (2005--2025) confirms that this 3-dimensional projection captures, on average, \textbf{67.5\%} of the total market variance, \ref{sec:results}.

It is crucial to note that while PCA imposes orthogonality (linear uncorrelatedness, $\mathbb{E}[F_i F_j] = 0$ for $i \neq j$), it does not imply statistical independence ($\mathbb{P}(F_i, F_j) \neq \mathbb{P}(F_i)\mathbb{P}(F_j)$), unless the underlying data is jointly Gaussian. Financial time series exhibit "stylized facts" such as volatility clustering and tail dependence \cite{Cont2001}, which imply significant non-linear coupling between uncorrelated modes. We hypothesize that this residual non-linear dependency manifests as the intrinsic \textit{curvature} of the manifold, distinguishing the market's true geometry from a flat Euclidean space.

\subsection{The Uniformization Conjecture: A Parity of Dimensions}
Our restriction to a 3-dimensional embedding is a deliberate topological choice rooted in the classification of surfaces. By modeling the market state as a trajectory on a 2-manifold embedded in $\mathbb{R}^3$, we utilize the \textbf{Uniformization Theorem} \cite{Abikoff1981, Poincare1907}. This fundamental result asserts that every simply connected Riemann surface is conformally equivalent to one of three canonical geometries:
\begin{itemize}
    \item \textbf{Spherical ($S^2$):} Positive curvature ($K>0$), implying compact, recurrent dynamics.
    \item \textbf{Euclidean ($\mathbb{R}^2$):} Zero curvature ($K=0$), implying flat, random-walk dynamics.
    \item \textbf{Hyperbolic ($H^2$):} Negative curvature ($K<0$), implying divergent or saddle-path dynamics.
\end{itemize}
We further include the Torus ($T^2$), which, although locally Euclidean, possesses a non-trivial topology ($S^1 \times S^1$) essential for capturing cyclical market phases.

Extending the model to include just one additional principal component (4 PCs) would require modeling a 3-manifold. Under the \textbf{Thurston-Perelman Geometrization Theorem} \cite{Thurston1997, Perelman2002}, the classification of 3-manifolds explodes in complexity, requiring decomposition into eight distinct geometries (e.g., $S^3, \mathbb{H}^3, \text{Nil}, \text{Sol}$). By limiting our scope to 3 PCs, we maintain a tractable geometric framework where the curvature sign provides clear, actionable market signals. However, we recognize that capturing the full spectrum of market variance may eventually require this higher-dimensional leap; consequently, extending our framework to the full eight Thurston geometries remains the primary objective of our future research.

\subsection{Curvature as an Alpha Signal}
The superior performance of the Geometry-Informed Model (GIM) over "flat" Euclidean baselines can be attributed to the correct specification of the transport dynamics. In a Euclidean framework (e.g., standard VAR), the expected path is a straight line in the tangent space, assuming zero acceleration.

However, if the market manifold is curved, a linear forecast systematically drifts off the manifold surface. The GIM corrects this by introducing a geometry-induced drift term derived from the \textit{mean curvature vector} $\mathbf{H}$:
\begin{equation}
    d\mathbf{X}_t = \mathbf{P}(\mathbf{X}_t) \circ d\mathbf{B}_t \approx \mathbf{P}(\mathbf{X}_t) d\mathbf{B}_t - \frac{1}{2} \mathbf{H}(\mathbf{X}_t) dt
\end{equation}
where $\mathbf{P}(\mathbf{X}_t)$ projects the noise onto the tangent space \cite{Hsu2002}. The term $-\frac{1}{2}\mathbf{H}$ acts as a "restoring force," pulling the trajectory along the manifold's geodesic. For example, in a spherical regime, this term enforces mean reversion naturally, whereas a flat model would predict a continued trend. By explicitly modeling this curvature, GIM anticipates non-linear turning points that linear models misinterpret as noise.

\subsection{Macroeconomic Intuition: The Latent Triad}
Beyond statistical variance and geometric tractability, the decision to target a 3-dimensional embedding finds a robust justification in macroeconomic theory. Financial asset prices are, in the long run, discounted claims on the real economy. Consequently, the dimensionality of the market's factor space should mirror the dimensionality of the underlying macroeconomic state vector.

Standard macroeconomic models, from the classical \textbf{IS-LM} framework \cite{Hicks1937} to modern \textbf{New Keynesian} Dynamic Stochastic General Equilibrium (DSGE) models \cite{Clarida1999, Woodford2003}, postulate that the economy is driven by three fundamental state variables:
\begin{enumerate}
    \item \textbf{Output Gap ($y_t$):} Represents real economic activity and the business cycle (The IS curve relation).
    \item \textbf{Inflation ($\pi_t$):} Represents the price level dynamics and aggregate supply constraints (The Phillips Curve).
    \item \textbf{Nominal Interest Rate ($i_t$):} Represents the policy stance and liquidity preference (The Taylor Rule or LM relation).
\end{enumerate}

This theoretical "triad" suggests that the essential dynamics of the economy evolve on a low-dimensional manifold embedded in a 3-dimensional phase space $(y_t, \pi_t, i_t)$. While a fourth dimension often identified as \textit{financial volatility} or risk premia plays a critical role, it frequently acts as a regime-switching parameter that alters the curvature of the space rather than an independent state variable of the same order.

Therefore, extracting the first three principal components is not merely a data compression technique; it is an attempt to recover the axes of this latent macroeconomic manifold. We hypothesize that our 3-dimensional eigenportfolio embedding $\mathbf{X}_t$ spans the same subspace as these fundamental macro-variables, allowing us to interpret the geometric "shape" of the market (e.g., the cyclical tori) as a direct manifestation of endogenous business cycles and policy feedback loops.

\section{Methodology}
\label{sec:method}

\paragraph{Roadmap.}
We first formalize Brownian motion constrained to embedded surfaces and derive explicit SDEs for the three geometries we use (Euclidean, spherical, hyperbolic), plus the torus via an intrinsic chart (Sec.~\ref{sec:sde_forms}). We then specify the logarithmic/exponential maps that move between the manifold and its tangent space (Sec.~\ref{sec:logexp}), followed by data-driven estimation of manifold parameters (sphere radius; torus radii; hyperboloid axes) directly from the observed 3D path (Sec.~\ref{subsec:manifold_param_estimation}). Next, we describe our curvature-based geometry inference with a torus topological check (Sec.~\ref{sec:curv_infer}), and the forecasting pipeline: chart $\to$ log map $\to$ tangent-space PCA \& time-series forecast $\to$ exponential map (Sec.~\ref{sec:forecasting}). We close with the explicit native-space baseline comparator and the translation from forecasts to volatility-scaled Profit and Loss (PnL) and portfolio construction (Sec.~\ref{subsec:real_fin_pipeline}). Simulation scenarios and additional implementation details are summarized in the Appendix to keep the section compact.

\section{Geometry-Constrained Stochastic Dynamics: Explicit Forms}

\subsection{Ambient formulation and curvature drift}
Let $M\subset\mathbb{R}^3$ be a $C^2$ embedded surface (2-manifold) with unit normal $n(x)$ at $x\in M$ and tangent projector
\[
P(x) \;=\; I - n(x)\,n(x)^\top \in\mathbb{R}^{3\times 3}.
\]
Brownian motion constrained to $M$ is the Stratonovich SDE (\cite{Hsu2002})
\begin{equation}\label{eq:strat}
dX_t \;=\; P(X_t)\circ dB_t,
\end{equation}
where $B_t$ is standard Brownian motion in $\mathbb{R}^3$. In Ito form this becomes (\cite{Hsu2002})
\begin{equation}\label{eq:ito}
dX_t \;=\; P(X_t)\,dB_t \;-\; \frac{1}{2}\,H(X_t)\,dt,
\end{equation}
where $H(x)$ is the \emph{mean curvature vector}, normal to $M$, that enforces the constraint (heuristically, $-\tfrac12 H$ pulls the path back onto $M$). For an \emph{implicit} surface $M=\{x: \phi(x)=0\}$ with $\nabla\phi(x)\neq 0$, we use
\[
n(x) \;=\; \frac{\nabla\phi(x)}{\|\nabla\phi(x)\|}, 
\qquad 
P(x) \;=\; I - n(x)n(x)^\top,
\]
and compute $H(x)$ either analytically (when available) or via the implementation used in our simulator (Appendix~\ref{app:sde_implementations}).

\subsection{Explicit SDEs by geometry}
\label{sec:sde_forms}

\paragraph{Euclidean ($\mathbb{R}^3$)(unconstained ambient).}
Trivial case (no constraint). With coordinates $X_t=(X_t^{(1)},X_t^{(2)},X_t^{(3)})$:
\begin{equation}\label{eq:euclid}
dX_t \;=\; dB_t.
\end{equation}

\paragraph{Sphere $S^2(R)$ (radius $R$).}
With $\phi(x)=\|x\|^2-R^2$, we have $n(x)=x/\|x\|$ and
\[
P(x)=I-\frac{x\,x^\top}{R^2}, 
\qquad 
H(x)=\frac{2}{R^2}\,x.
\]
Hence the Ito SDE on $S^2(R)$ is (\cite{Hsu2002})
\begin{equation}\label{eq:sphere_sde}
dX_t \;=\; \Big(I-\frac{X_t X_t^\top}{R^2}\Big)\,dB_t \;-\; \frac{1}{R^2}X_t\,dt.
\end{equation}
For $R=1$ this reduces to $dX_t = (I-X_tX_t^\top)dB_t - X_t\,dt$.

\paragraph{Torus $T^2(R,r)$ (major radius $R$, minor radius $r$).}
We work both in implicit embedding and in intrinsic angles.
\begin{itemize}\itemsep 0.25em
\item \emph{Implicit embedding:} $\displaystyle \phi(x)=\big(R-\sqrt{x_1^2+x_2^2}\big)^2 + x_3^2 - r^2=0$. Then
\[
n(x)=\frac{\nabla\phi(x)}{\|\nabla\phi(x)\|},\qquad P(x)=I-n(x)n(x)^\top,
\]
and we evolve \eqref{eq:ito} with this projector and the corresponding $H(x)$ (Appendix~\ref{app:sde_implementations}).
\item \emph{Intrinsic chart $(\theta,\varphi)\in[0,2\pi)^2$:}
\[
\Psi(\theta,\varphi)=
\begin{pmatrix}
(R+r\cos\varphi)\cos\theta\\[0.2em]
(R+r\cos\varphi)\sin\theta\\[0.2em]
r\sin\varphi
\end{pmatrix}.
\]
The metric is diagonal: $g_{\theta\theta}=(R+r\cos\varphi)^2$, $g_{\varphi\varphi}=r^2$,\cite{doCarmoRiemannian}, hence $g^{\theta\theta}=(R+r\cos\varphi)^{-2}$, $g^{\varphi\varphi}=r^{-2}$ and $\sqrt{|g|}=r(R+r\cos\varphi)$. The Ito SDE realizing generator $\tfrac12\Delta_g$ is
\begin{align}
d\theta_t &= \frac{1}{R+r\cos\varphi_t}\,dW_t^{(1)}, \label{eq:torus_theta}\\
d\varphi_t &= \frac{1}{r}\,dW_t^{(2)} \;-\; \frac{\sin\varphi_t}{2r\,(R+r\cos\varphi_t)}\,dt. \label{eq:torus_phi}
\end{align}
Cartesian positions follow by $X_t=\Psi(\theta_t,\varphi_t)$.
\end{itemize}

\paragraph{Hyperbolic $H^2$ (hyperboloid model).}
Use parameters $(u,v)\in\mathbb{R}\times[0,2\pi)$ and constants $a>0$, $c>0$:
\[
\Phi(u,v)=\begin{pmatrix}
a\cosh u\cos v\\ a\cosh u\sin v\\ c\sinh u
\end{pmatrix}.
\]
Then $g_{uu}=E(u)=a^2\sinh^2 u + c^2\cosh^2 u$, $g_{vv}=G(u)=a^2\cosh^2 u$,\cite{doCarmoRiemannian}, so $g^{uu}=1/E$, $g^{vv}=1/G$, $\sqrt{|g|}=\sqrt{E(u)G(u)}=a\cosh u\,\sqrt{E(u)}$. The Ito SDE (generator $\tfrac12\Delta_g$) reads
\begin{align}
d u_t &= \frac{1}{\sqrt{E(u_t)}}\,dW_t^{(1)} \;+\; \frac{1}{2}\,\partial_u\!\left[\ln\!\big(\sqrt{|g|}\,g^{uu}\big)\right]_{u=u_t}\,dt \nonumber\\
      &= \frac{1}{\sqrt{E(u_t)}}\,dW_t^{(1)} \;+\; \frac{1}{2}\left(\tanh u_t - \frac{E'(u_t)}{2E(u_t)}\right)dt, \label{eq:hyper_u}\\
d v_t &= \frac{1}{\sqrt{G(u_t)}}\,dW_t^{(2)} \;=\; \frac{1}{a\cosh u_t}\,dW_t^{(2)}, \label{eq:hyper_v}
\end{align}
with $E'(u)=2(a^2+c^2)\sinh u\cosh u$. Cartesian positions are $X_t=\Phi(u_t,v_t)$. 
\vspace{0.25em}

\noindent\textbf{Remark (ambient implementation).} We also implement \eqref{eq:ito} directly in $\mathbb{R}^3$ using $P(x)$ and a closed-form mean-curvature drift $H(x)$ for each implicit surface; formulas above and the implementation are equivalent modulo time discretization (Appendix~\ref{app:sde_implementations}).

\subsection{Nonlinear forecasting via machine learning regressors}
\label{subsec:nonlinear_forecasting}

Beyond the linear Vector AutoRegression (VAR) model, we also tested two nonlinear regressors—\emph{Random Forests (RF)} and \emph{Gaussian Process Regression (GP)}—to assess whether local nonlinearities in the tangent-space coordinates could enhance predictive accuracy. 
Both approaches share the same input representation, where the target variable $y_t$ (one coordinate of the tangent vector series) is regressed on its $L$ most recent lags:
\begin{equation}
\mathcal{D} = \{(\mathbf{x}_t, y_t)\}_{t=L}^{T-1}, 
\qquad 
\mathbf{x}_t = [y_{t-1}, y_{t-2}, \dots, y_{t-L}]^\top.
\end{equation}
At each step, the models are trained on $\mathcal{D}$ and used to predict $\widehat{y}_{T+1} = f(\mathbf{x}_T)$, where $f(\cdot)$ denotes the learned regression mapping.

\paragraph{Random Forest regression.}
A Random Forest \citep{breiman2001random} constructs an ensemble of $B$ regression trees $\{f_b(\cdot)\}_{b=1}^{B}$, each trained on a bootstrap resample of $\mathcal{D}$ and a random subset of predictors. 
The forecast corresponds to the ensemble mean:
\begin{equation}
\widehat{y}_{T+1}^{(\mathrm{RF})} = \frac{1}{B}\sum_{b=1}^{B} f_b(\mathbf{x}_T),
\end{equation}
where each $f_b$ partitions the lagged feature space $\mathbb{R}^L$ into piecewise-constant regions and averages the training observations within the corresponding leaf.
RF models approximate nonlinear relationships and capture variable interactions without assuming parametric structure. 
In our implementation, we used $B=100$ trees and a memory length of $L=5$ lags for all tangent coordinates $(x_t, y_t, z_t)$.

\paragraph{Gaussian Process regression.}
The Gaussian Process (GP) model treats the regression function $f(\cdot)$ as a random function drawn from a Gaussian process prior:
\begin{equation}
f(\cdot) \sim \mathcal{GP}(m(\cdot), k(\cdot,\cdot)),
\end{equation}
with mean function $m(\cdot)$ (set to zero) and covariance kernel $k(\mathbf{x},\mathbf{x}')$ encoding smoothness and correlation structure. 
Given training data $\mathcal{D}$, the posterior predictive distribution at a new input $\mathbf{x}_T$ is Gaussian:
\begin{equation}
p(y_{T+1}\,|\,\mathbf{x}_T,\mathcal{D})
= 
\mathcal{N}\big(\mathbf{k}_T^\top(K+\sigma_n^2 I)^{-1}\mathbf{y},\;
k(\mathbf{x}_T,\mathbf{x}_T) - \mathbf{k}_T^\top(K+\sigma_n^2 I)^{-1}\mathbf{k}_T\big),
\end{equation}
where $K$ is the kernel matrix with $[K]_{ij}=k(\mathbf{x}_i,\mathbf{x}_j)$, and $\mathbf{k}_T = [k(\mathbf{x}_1,\mathbf{x}_T),\dots,k(\mathbf{x}_N,\mathbf{x}_T)]^\top$. 
The mean of this posterior gives the forecast:
\begin{equation}
\widehat{y}_{T+1}^{(\mathrm{GP})} = \mathbf{k}_T^\top (K+\sigma_n^2 I)^{-1}\mathbf{y}.
\end{equation}
We used a Matérn kernel with an additive constant term,
\begin{equation}
k(\mathbf{x},\mathbf{x}') = k_{\mathrm{Matern}}(\mathbf{x},\mathbf{x}') + k_{\mathrm{const}}(\mathbf{x},\mathbf{x}'),
\end{equation}
which balances smooth local trends with global level shifts, providing flexibility for financial data where both slow drifts and abrupt changes can occur.

\paragraph{Forecast integration.}
In the forecasting pipeline, both RF and GP models operate on the tangent-space coordinates and provide one-step-ahead predictions $\widehat{v}_{t+1}$, which are then lifted back to the manifold via the exponential map,
\begin{equation}
\widehat{x}_{t+1} = \exp_{\widehat{\mu}}(\widehat{v}_{t+1}),
\end{equation}
thus preserving the geometric structure of the original process. For our analysis, we used a rolling window framework for both ML methods, using 25 observations (trading days) for training and predicting one-step ahead points. 

\subsection{Logarithmic and exponential mappings}
\label{sec:logexp}

Let $T_xM$ denote the tangent plane at $x\in M$.\cite{Pennec2018,doCarmoRiemannian}
\paragraph{Sphere $S^2(R)$.} For unit vectors $\mu,p\in S^2(R)$ define $\theta=\arccos\!\big(\tfrac{\mu\cdot p}{R^2}\big)\in[0,\pi]$. Then
\begin{align}
\log_{\mu}(p)
&=
\begin{cases}
\mathbf{0}, & p=\mu,\\[0.35em]
\dfrac{\theta}{\sin\theta}\;\Pi_\mu\,p, & p\neq \mu,
\end{cases}
\label{eq:log_sphere}
\\[0.25em]
\exp_\mu(v)
&=\cos\!\big(\|v\|/R\big)\,\mu
\;+\;
R\,\sin\!\big(\|v\|/R\big)\,\dfrac{v}{\|v\|},
\qquad v\in T_\mu S^2(R).
\label{eq:exp_sphere}
\end{align}
\noindent See \cite{Pennec2018,doCarmoRiemannian} for closed forms.
\paragraph{Torus $T^2(R,r)$ (angle chart).\cite{doCarmoCurves}} Use angles $(\theta,\varphi)\in[0,2\pi)^2$. For base point $(\theta_0,\varphi_0)$ and point $(\theta,\varphi)$,
\begin{align}
\log_{(\theta_0,\varphi_0)}(\theta,\varphi) 
&= \big( \mathrm{wrap}(\theta-\theta_0),\; \mathrm{wrap}(\varphi-\varphi_0)\big), \label{eq:log_torus}\\
\exp_{(\theta_0,\varphi_0)}(\Delta\theta,\Delta\varphi) 
&= \big( \theta_0+\Delta\theta,\; \varphi_0+\Delta\varphi\big)\bmod 2\pi, \label{eq:exp_torus}
\end{align}
where $\mathrm{wrap}(\alpha)=((\alpha+\pi)\bmod 2\pi)-\pi$ is the $[-\pi,\pi)$ minimizer. Mapping to $\mathbb{R}^3$ uses $\Psi$ above.

\paragraph{Hyperbolic (hyperboloid chart).\cite{doCarmoRiemannian}} In coordinates $(u,v)$ with metric $\mathrm{diag}(E(u),G(u))$,
\begin{align}
\log_{(u_0,v_0)}(u,v) &= \big(u-u_0,\; \mathrm{wrap}(v-v_0)\big), \label{eq:log_hyper}\\
\exp_{(u_0,v_0)}(\Delta u,\Delta v) &= \big( u_0+\Delta u,\; v_0+\Delta v \bmod 2\pi\big), \label{eq:exp_hyper}
\end{align}
then lift to $\mathbb{R}^3$ via $\Phi$. For small moves these coincide with the Riemannian log/exp in these orthogonal coordinates.

\paragraph{Karcher mean and tangent PCA.\cite{Karcher1977,Pennec2018}} Given points $\{x_i\}\subset M$, we compute the intrinsic (Karcher) mean $\mu^\star$ by iterating $\mu\leftarrow \exp_\mu\!\big( \tfrac{1}{N}\sum_i \log_\mu(x_i)\big)$ until convergence. We then project data to $T_{\mu^\star}M$ with $\log_{\mu^\star}(x_i)$ and perform PCA there (this is the “tangent-PCA” used in our forecasting pipeline).

\subsection{Data-driven estimation of manifold parameters}
\label{subsec:manifold_param_estimation}

A key ingredient of our pipeline is that the geometry is not treated as fixed; instead, its \emph{intrinsic parameters} are inferred on the fly from the observed 3D path $\{x_t=(x_t,y_t,z_t)\}_{t\le t^\star}$. This subsection formalizes the estimation steps we implement for the sphere, torus and one–sheeted hyperboloid, matching the code used in our experiments.\cite{CazalsPouget2005,CohenSteiner2006}

\paragraph{Notation.} Let $M(\vartheta)$ denote a parametric surface embedded in $\mathbb{R}^3$ with parameter vector $\vartheta$ (e.g., radius $R$ on the sphere; major/minor radii $(R,r)$ on the torus; semi–axes $(a,b,c)$ on the hyperboloid). Given a stream $\{x_t\}$ up to time $t^\star$, we estimate $\widehat{\vartheta}_{t^\star}$ and then work in the corresponding chart for mapping, forecasting and lifting. Throughout, angles are unwrapped \emph{mod} $2\pi$ using the shortest–-arc convention to preserve continuity in the tangent space (see~\eqref{eq:shortest_arc} below).

\subsubsection*{Sphere $S^2(R)$: Karcher mean and radius}
For spherical segments, we first compute a \emph{Karcher (Fréchet) mean} $\widehat{\mu}\in S^2$ of the points $\{x_t/\|x_t\|\}$ by iterating $\mu \leftarrow \exp_\mu\!\big(\tfrac{1}{n}\sum_{t}\log_\mu(x_t)\big)$ until convergence;\cite{Karcher1977}

An estimate of the radius follows from
\[
\widehat{R}_{t^\star}
\;=\;
\operatorname*{arg\,min}_{R>0}\,
\sum_{t\le t^\star}\big(\|x_t\|-R\big)^2
\;=\;
\frac{1}{t^\star}\sum_{t\le t^\star}\|x_t\|.
\]
This $\widehat{R}$ is needed since we wish to keep track of the physical scale; 

\subsubsection*{Torus $T^2(R,r)$: method–-of-–moments from toroidal coordinates}
We convert Cartesian observations to toroidal angles $(\theta_t,\phi_t)$ and the auxiliary radial quantity $\rho_t=\sqrt{x_t^2+y_t^2}$ : a standard torus (major radius $R$, minor radius $r$) in the \emph{Reinhardt} parameterization is
\[
x=(R+r\cos\phi)\cos\theta,\quad
y=(R+r\cos\phi)\sin\theta,\quad
z=r\sin\phi.
\]
From $\rho_t = \sqrt{x_t^2+y_t^2} = R + r\cos\phi_t$ it follows that
\(
\mathbb{E}[\rho_t] = R,\quad
\operatorname{Var}(\rho_t)=\tfrac{r^2}{2}\ \text{if } \phi_t\sim \text{Unif}.
\)
We adopt a simple method of moments on the \emph{minor angle} and its cosine (normalized units):
\begin{equation}
\widehat{R} \;=\; \frac{1}{t^\star}\sum_{t\le t^\star}\cos\phi_t,
\qquad
\widehat{r}\;=\;
\sqrt{\frac{1}{t^\star}\sum_{t\le t^\star}\big(\cos\phi_t-\widehat{R}\big)^2},
\label{eq:torus_MoM}
\end{equation}

In practice, for stability we use $\widehat{R}$ from~\eqref{eq:torus_MoM} while the \emph{instantaneous} tube radius is proxied by the latest $\rho_{t^\star}$, which helps track slow deformations of the tube thickness in non–stationary segments. \cite{doCarmoCurves}

Tangent space steps use the $2\pi$ wrapped \emph{shortest arc} differences
\begin{equation}
\Delta\theta_t \;=\; ((\theta_t-\theta_0+\pi)\bmod 2\pi)-\pi,
\qquad
\Delta\phi_t \;=\; ((\phi_t-\phi_0+\pi)\bmod 2\pi)-\pi,
\label{eq:shortest_arc}
\end{equation}

\subsubsection*{One sheeted hyperboloid $x^2/a^2+y^2/b^2-z^2/c^2=1$: nonlinear Least Squares}
For hyperbolic patches we assume the one–sheeted quadratic model
\[
F(x,y,z;a,b,c)\;=\;\frac{x^2}{a^2}+\frac{y^2}{b^2}-\frac{z^2}{c^2}-1 \;=\;0.
\]
Given $\{x_t\}$, we fit $(a,b,c)$ by nonlinear least squares
\begin{equation}
(\widehat{a},\widehat{b},\widehat{c})
\;\in\;\arg\min_{a,b,c>0}\ \sum_{t\le t^\star} F(x_t,y_t,z_t;a,b,c)^2,
\label{eq:hyper_fit}
\end{equation}
just before transforming the stream to hyperbolic coordinates and entering the log, forecast, exp cycle

The coordinate chart uses $(u,v)$ with
\(
x=a\cosh u\cos v,\ y=b\cosh u\sin v,\ z=c\sinh u
\),
and the forward/backward conversions are handled by implementing the numerically stable choice $u=\operatorname{arcsinh}(z/c)$ and $v=\operatorname{atan2}(y/b,x/a)$). 

On this chart, the log map again uses shortest arc wrapping on $v$ and simple differencing on $u$; the exp map re–adds those increments around the base point. \cite{doCarmoRiemannian}

\subsubsection*{Chart transition inside the 'motion mixing' framework}
At each step $t^\star$, the pipeline (i) picks/updates the active geometry $M(\widehat{\vartheta}_{t^\star})$, (ii) \emph{re–charts} the recent window to the intrinsic coordinates of $M$, (iii) applies \emph{tangent space} PCA/forecasting, and (iv) lifts the predicted tangent vector back to $\mathbb{R}^3$ through the geometry exp map and its embedding. 

Concretely:

\noindent\textbf{Sphere (S).}\;
\(
x_t \;\mapsto\; \tilde x_t:=\frac{x_t}{\|x_t\|}\in S^2(R)
\;\xrightarrow{\ \log_{\widehat\mu}\ }\;
v_t\in T_{\widehat\mu}S^2(R)
\;\xrightarrow{\ \text{PGA + forecast}\ }\;
\widehat v_{t+1}
\;\xrightarrow{\ \exp_{\widehat\mu}\ }\;
\widehat x_{t+1}.
\)

\medskip
\noindent\textbf{Torus (T).}\;
\(
x_t \;\mapsto\; (\theta_t,\phi_t)
\;\xrightarrow{\ \log_{(\theta_0,\phi_0)}\ }\;
v_t
\;\xrightarrow{\ \text{PGA + forecast}\ }\;
\widehat v_{t+1}
\;\xrightarrow{\ \exp_{(\theta_0,\phi_0)}\ }\;
(\widehat\theta,\widehat\phi)
\;\xrightarrow{\ \Psi_{(\widehat R,\widehat r)}\ }\;
\widehat x_{t+1},
\)
where \(\Psi_{(R,r)}:S^1\times S^1\to\mathbb{R}^3\) is the standard torus embedding.

\medskip
\noindent\textbf{Hyperbolic (H).}\;
\(
x_t \;\mapsto\; (u_t,v_t)
\;\xrightarrow{\ \log_{(u_0,v_0)}\ }\;
w_t
\;\xrightarrow{\ \text{PGA + forecast}\ }\;
\widehat w_{t+1}
\;\xrightarrow{\ \exp_{(u_0,v_0)}\ }\;
(\widehat u,\widehat v)
\;\xrightarrow{\ \Phi_{(\widehat a,\widehat b,\widehat c)}\ }\;
\widehat x_{t+1},
\)
where \(\Phi_{(a,b,c)}\) embeds the hyperbolic chart (e.g., hyperboloid model) into \(\mathbb{R}^3\).

By explicitly re–estimating $(\widehat R)$ or $(\widehat R,r)$ or $(\widehat a,\widehat b,\widehat c)$ at each step (or on an expanding/rolling schedule), we avoid confusing the \emph{learned geometry} with the auxiliary linear reduction used inside the tangent space,\cite{Pennec2018}. The code performs these updates right before building the log–-tangent cloud and forecasting the principal coordinates. 

\subsubsection*{Numerical considerations}
\begin{itemize}
\item \textbf{Angle wrapping.} All angular differences use the shortest–arc rule~\eqref{eq:shortest_arc} to keep tangent vectors small and numerically stable on compact directions (torus $S^1{\times}S^1$ and the angular coordinate $v$ on the hyperboloid). 
\item \textbf{Stability at small steps.} Spherical log/exp guard against $\sin\theta\!\approx\!0$ and $\|v\|\!\approx\!0$ (returning zeros or the base point), preventing blow–ups when points are nearly aligned. 
\item \textbf{Robust lifting.} The torus and hyperboloid use explicit closed–form embeddings for \emph{lifting} the predicted coordinates back to $\mathbb{R}^3$, ensuring that $\widehat{x}_{t+1}$ lies on $M(\widehat\vartheta_{t^\star})$ by construction. 
\end{itemize}

\subsection{Local Gaussian Curvature Based Geometry Inference}
\label{sec:curv_infer}

\begin{figure}[H]
\centering
% Row 1
\begin{minipage}{0.24\linewidth}\centering
\includegraphics[width=\linewidth]{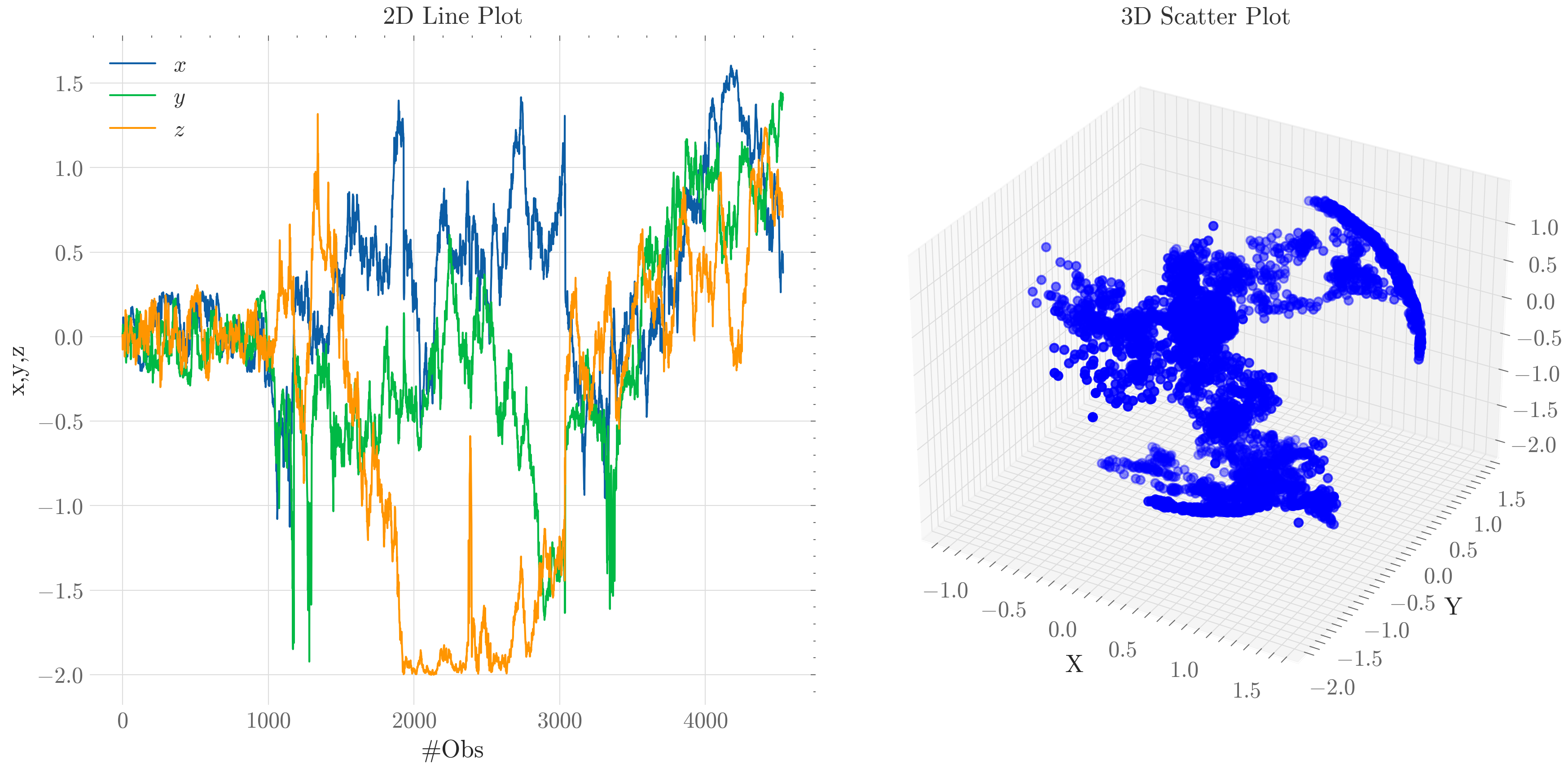}\\
{\small (a) Scenario 1}
\end{minipage}\hfill
\begin{minipage}{0.24\linewidth}\centering
\includegraphics[width=\linewidth]{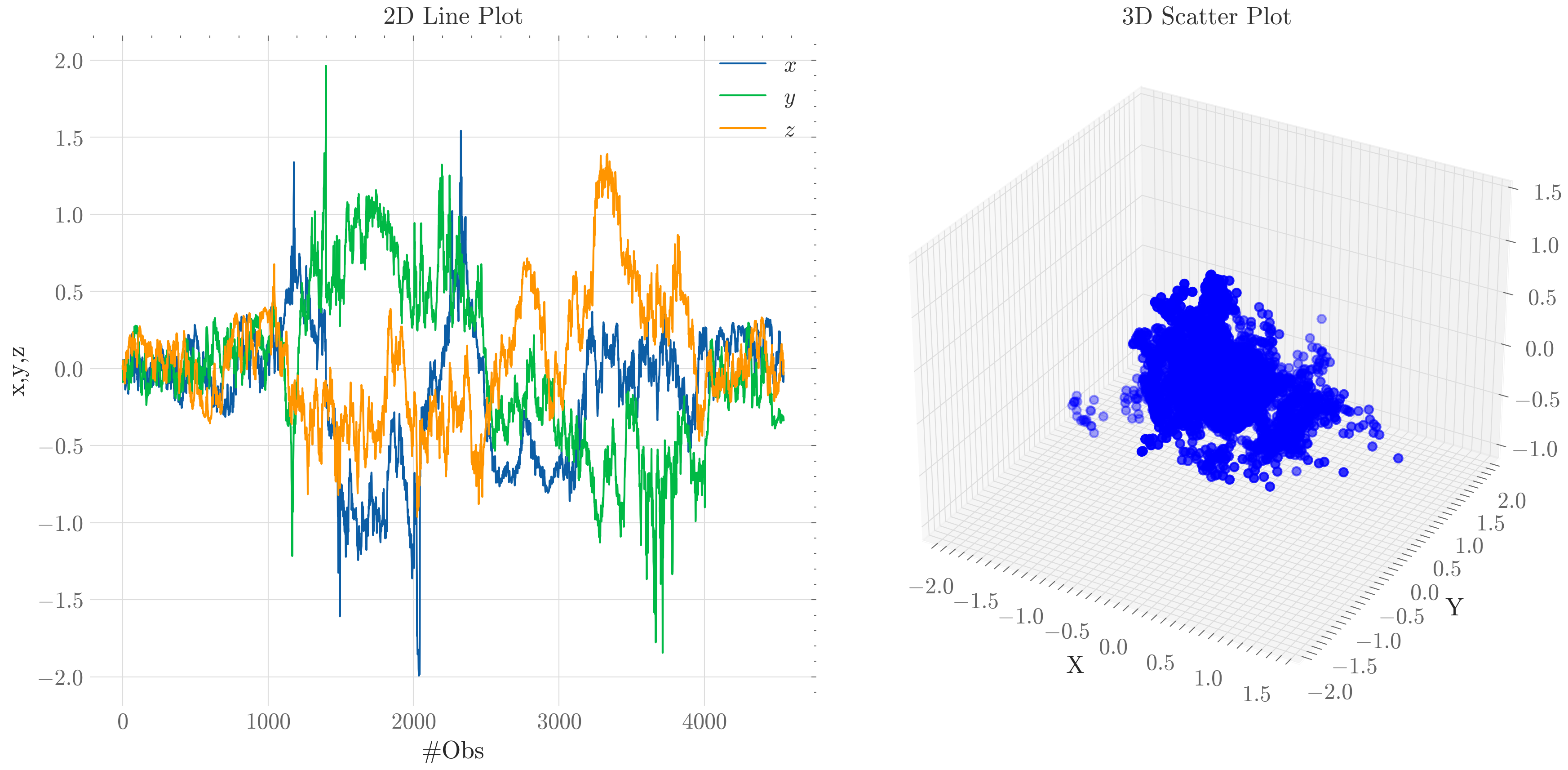}\\
{\small (b) Scenario 2}
\end{minipage}\hfill
\begin{minipage}{0.24\linewidth}\centering
\includegraphics[width=\linewidth]{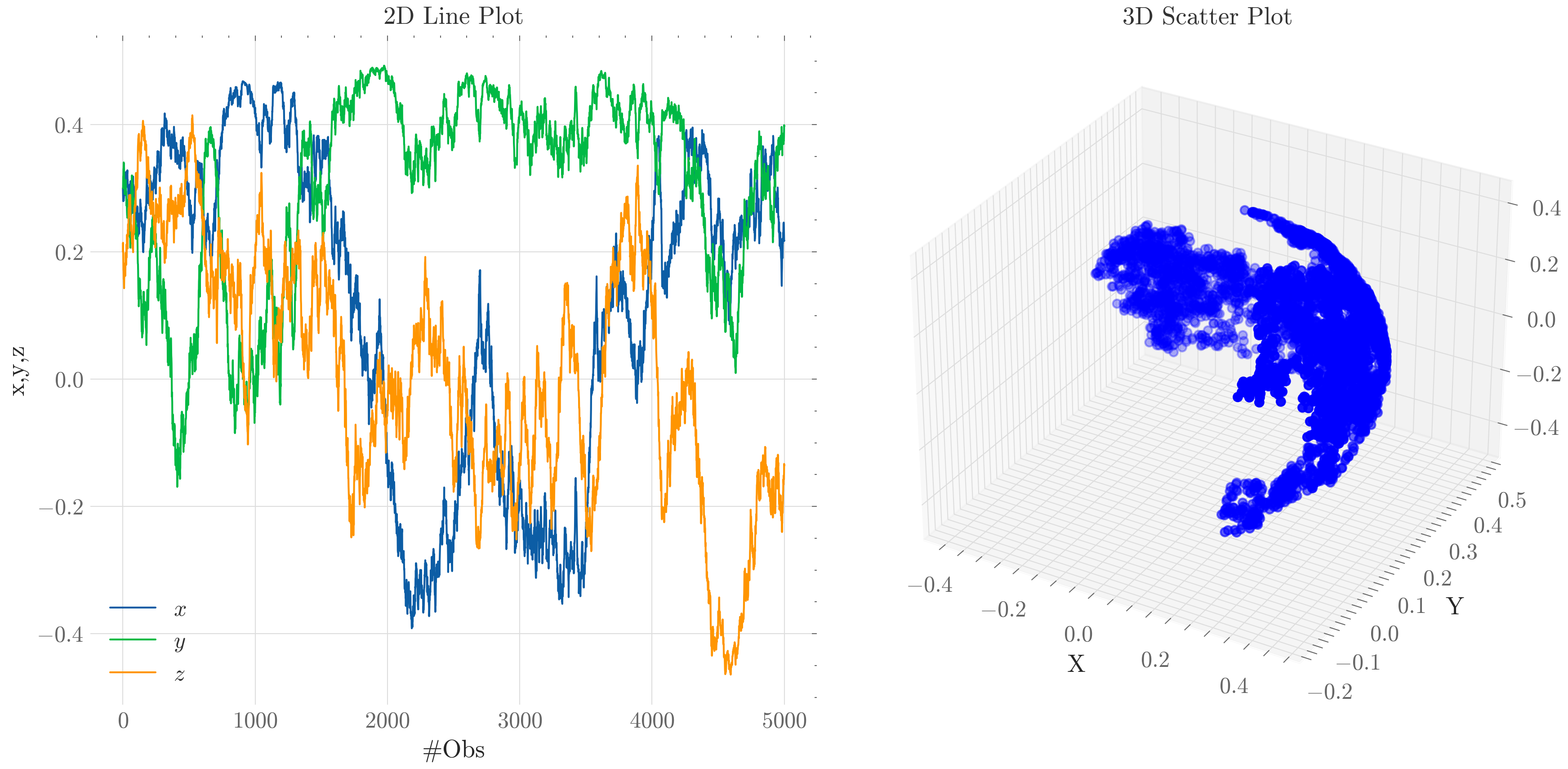}\\
{\small (c) Scenario 3}
\end{minipage}\hfill
\begin{minipage}{0.24\linewidth}\centering
\includegraphics[width=\linewidth]{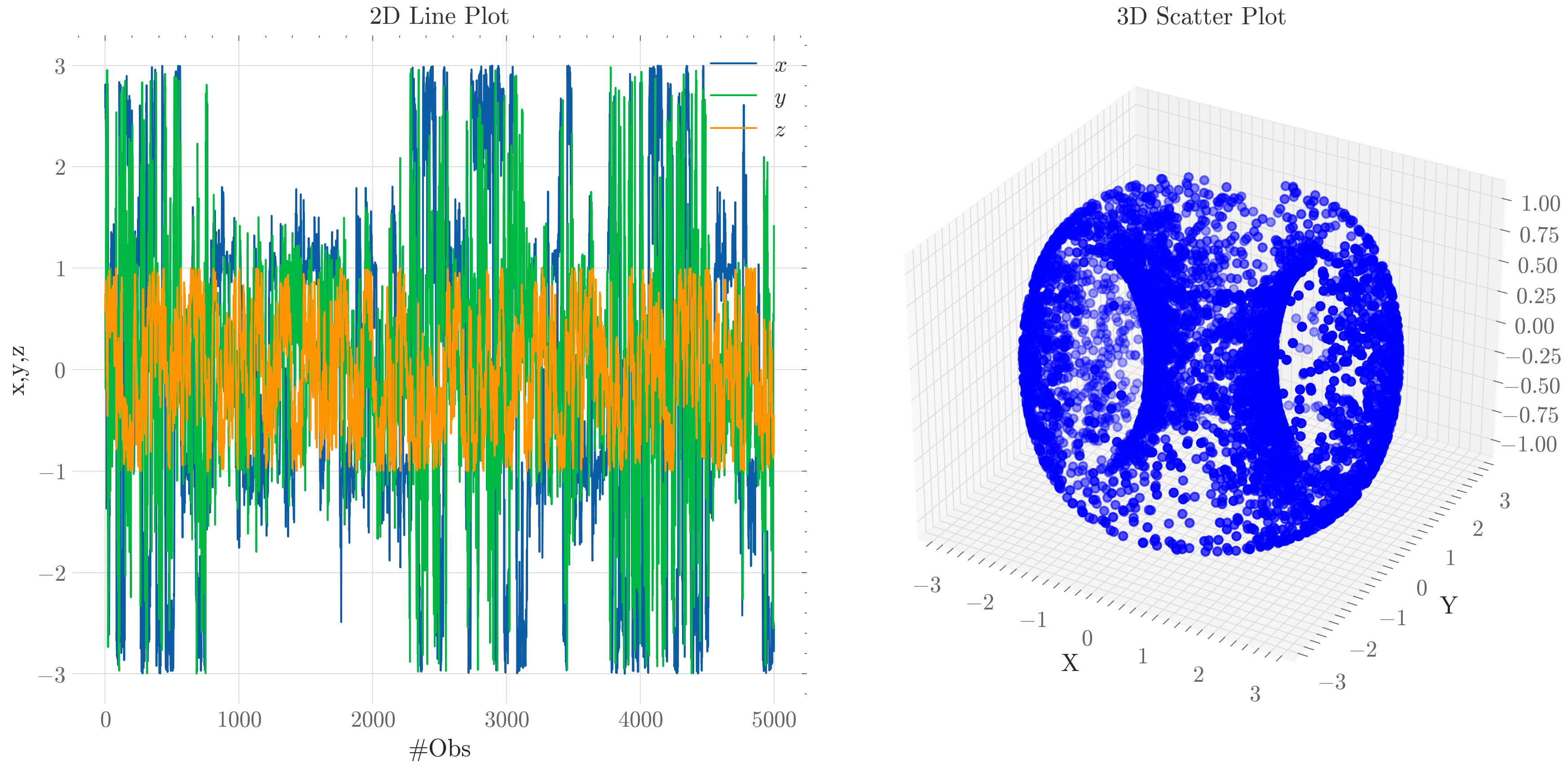}\\
{\small (d) Scenario 4}
\end{minipage}

\vspace{0.6em}
% Row 2
\begin{minipage}{0.24\linewidth}\centering
\includegraphics[width=\linewidth]{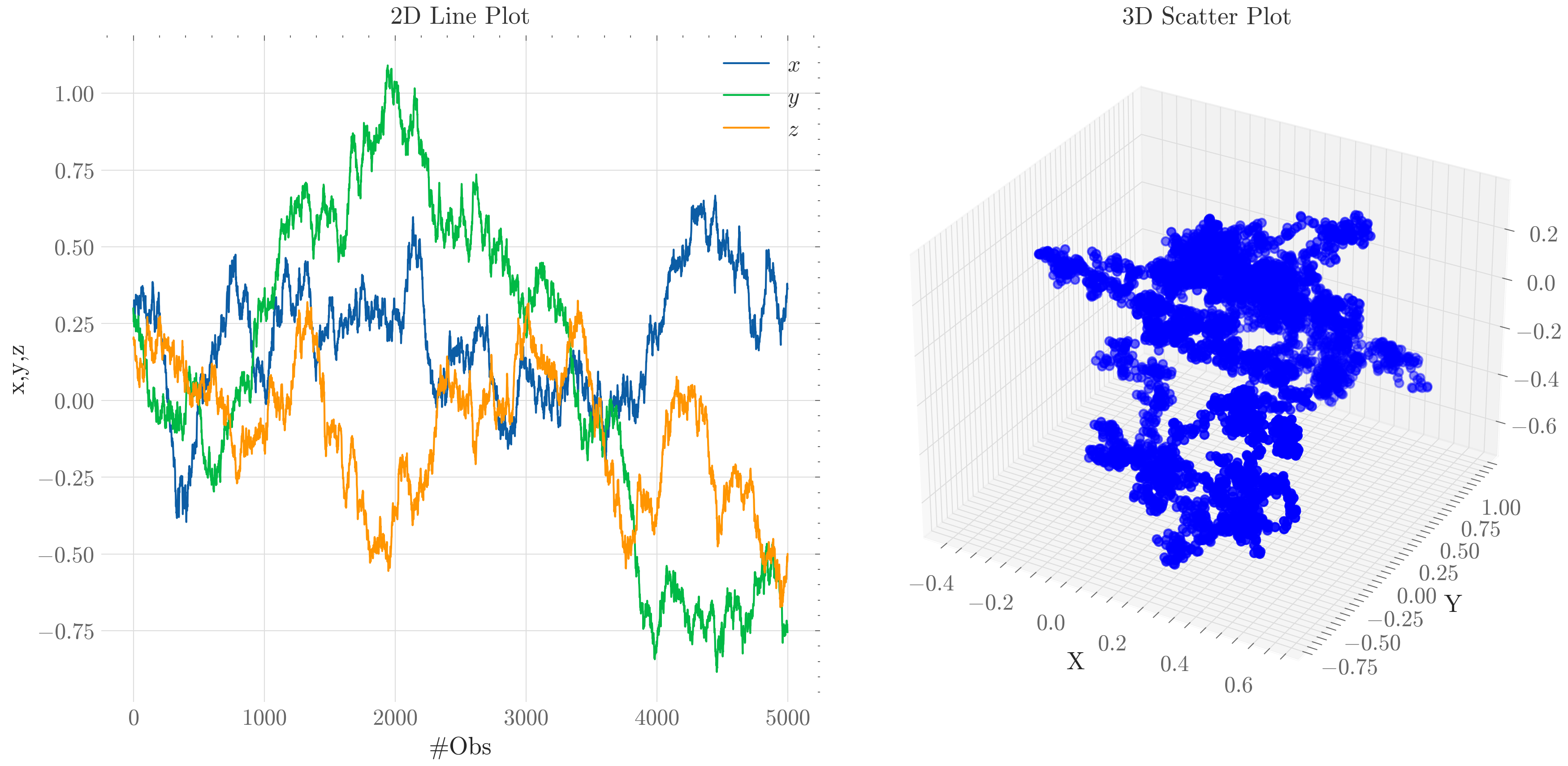}\\
{\small (e) Scenario 5}
\end{minipage}\hfill
\begin{minipage}{0.24\linewidth}\centering
\includegraphics[width=\linewidth]{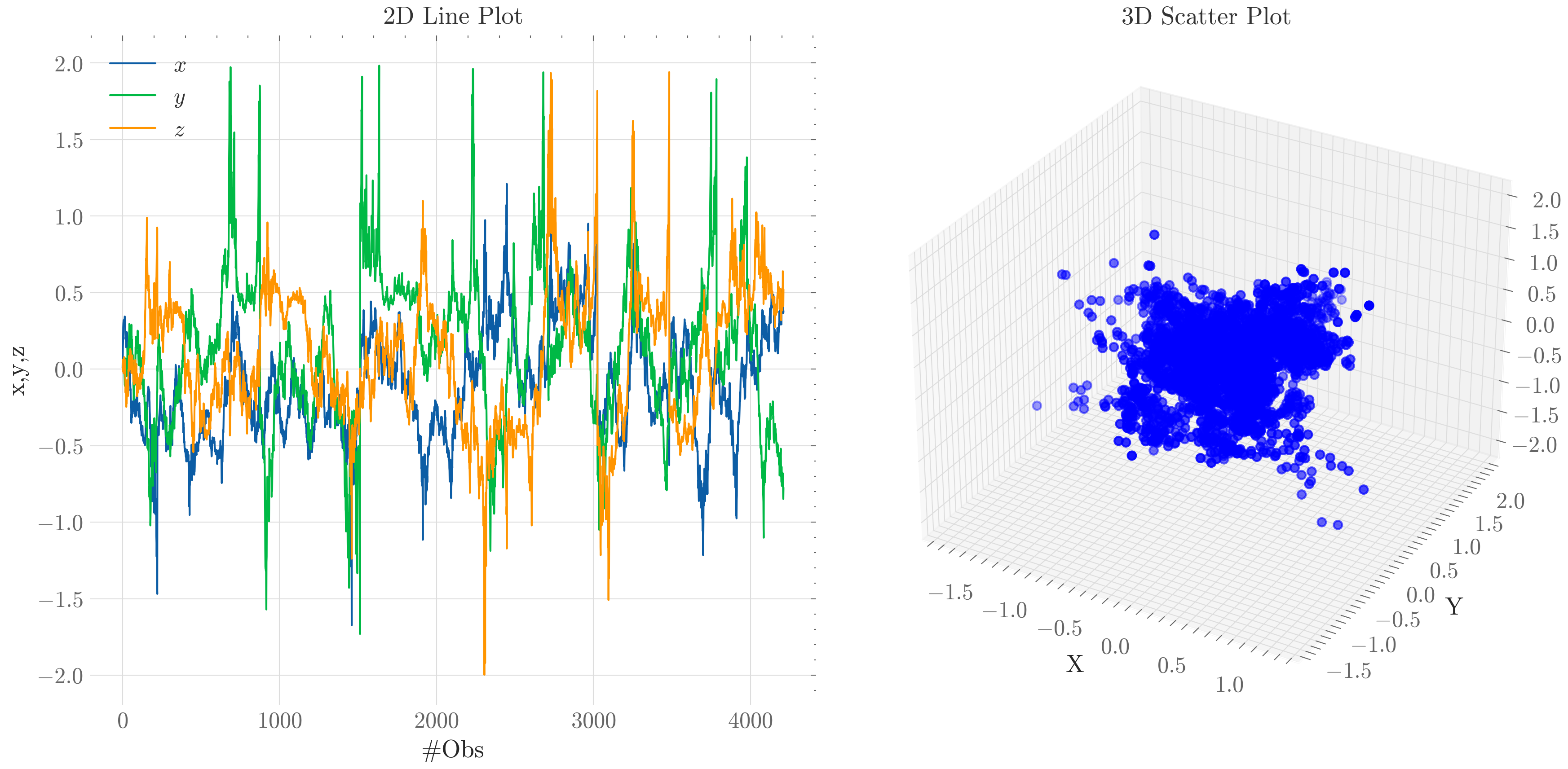}\\
{\small (f) Scenario 6}
\end{minipage}\hfill
\begin{minipage}{0.24\linewidth}\centering
\includegraphics[width=\linewidth]{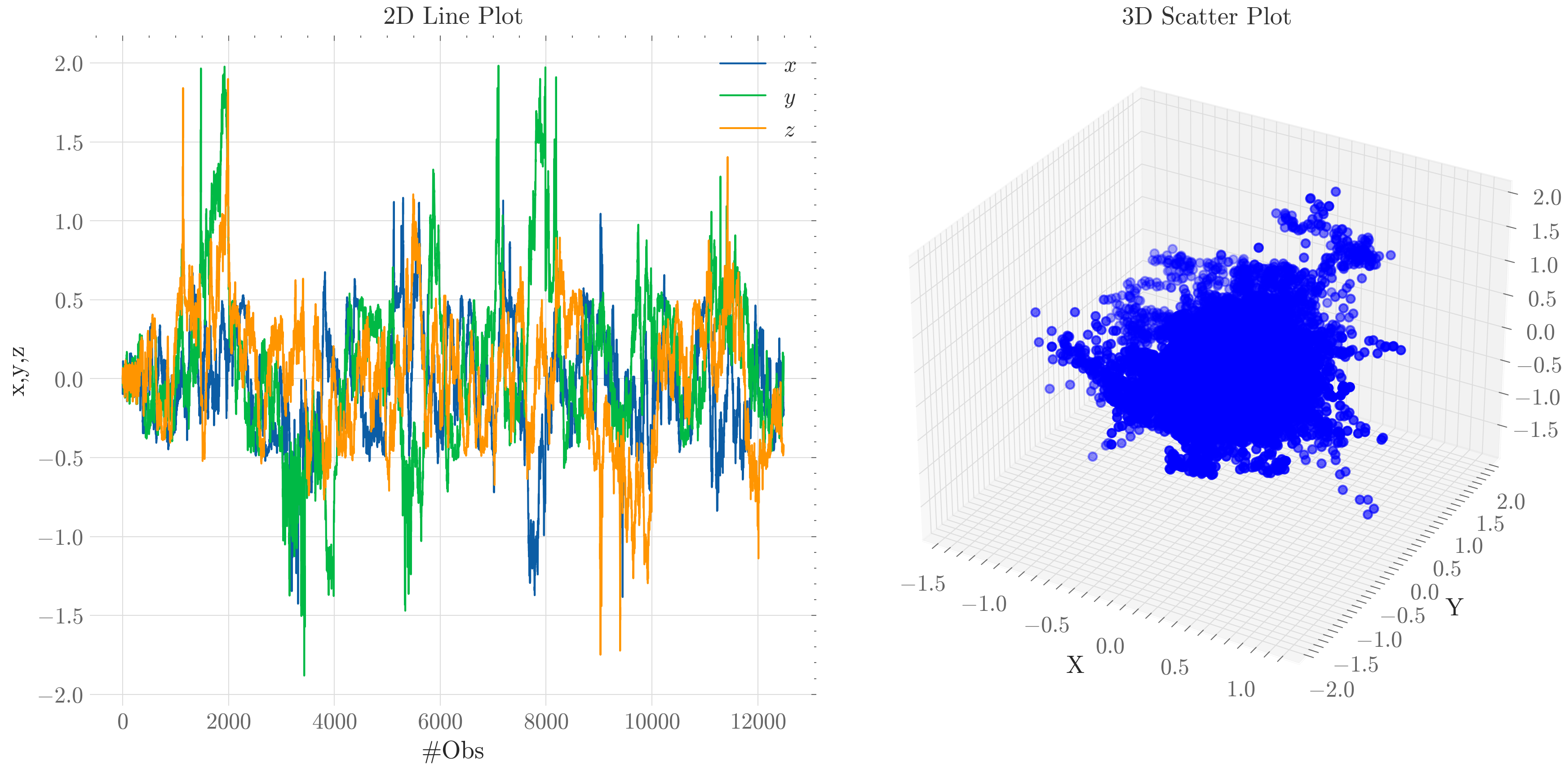}\\
{\small (g) Scenario 7}
\end{minipage}\hfill
\begin{minipage}{0.24\linewidth}\centering
\includegraphics[width=\linewidth]{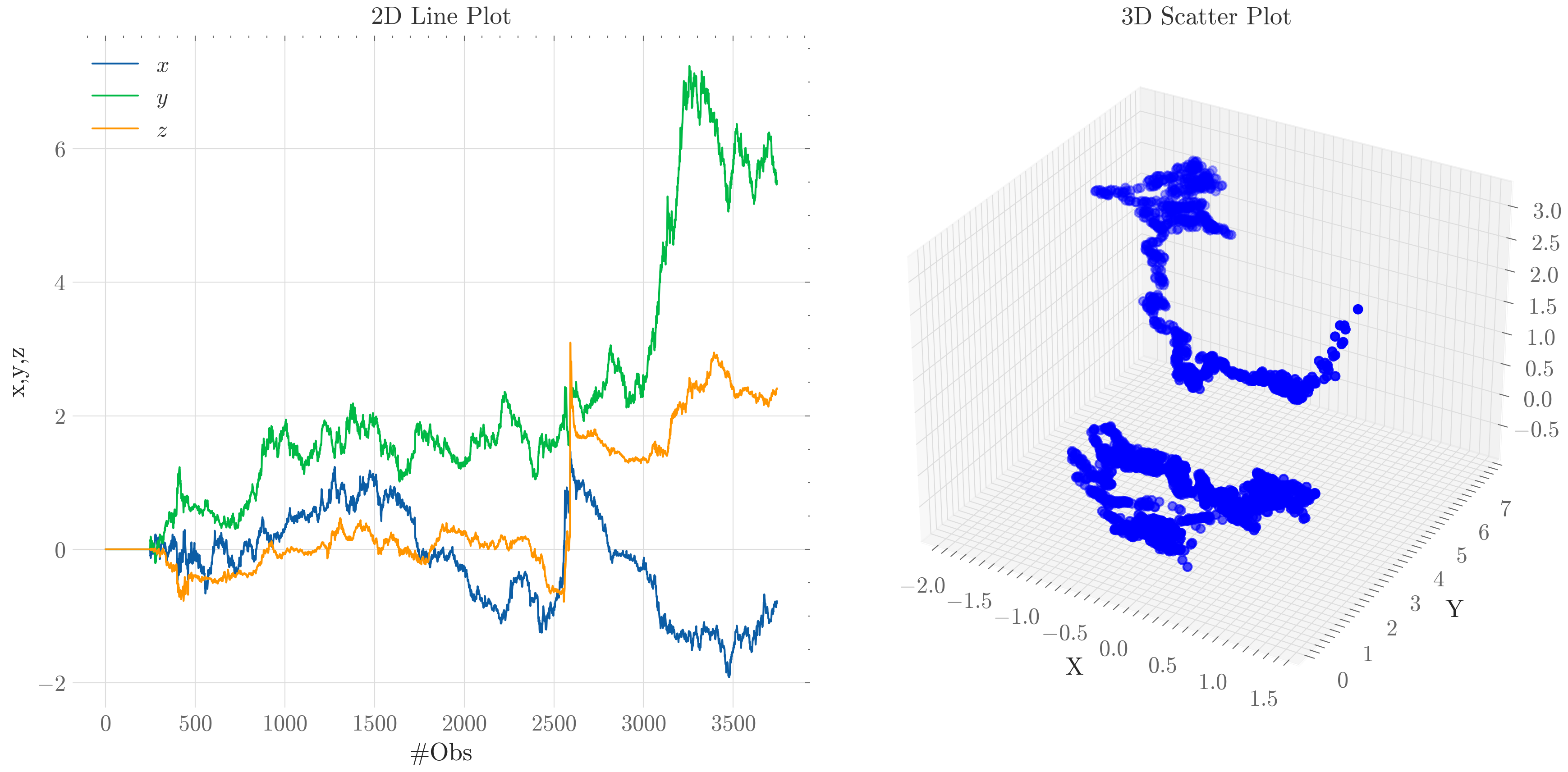}\\
{\small (h) Financial data 3D PCA embedding}
\end{minipage}

\caption{Simulated Brownian-motion scenarios: time-series panels (a–g) and the corresponding 3D PCA embedding (h).}
\label{fig:bm_multifigure_minipage}
\end{figure}

We infer the latent geometry directly from the data via a \emph{local differential–geometric fit} combined with \emph{topological validation}. The pipeline operates on a 3D trajectory
$X_t=(x_t,y_t,z_t)\in\mathbb{R}^3$ obtained either from simulation or from real data after a 3D embedding (e.g., expanding-window PCA of returns). We show the dynamics of the simulations and from the real data in \ref{fig:bm_multifigure_minipage}.

At each time $t$ we consider a window $\mathcal{W}_t=\{X_s:s\in[t_0,t]\}$ ( expanding) with $|\mathcal{W}_t|\ge m_0$ points and compute a local quadratic Monge patch fit,\cite{CazalsPouget2005,CohenSteiner2006},
\begin{equation}
z \;=\; a x^2 + b x y + c y^2 + d x + e y + f \quad \text{over } \mathcal{W}_t,
\label{eq:quadfit}
\end{equation}
by weighted least squares. Writing $A=\big[x^2,\,xy,\,y^2,\,x,\,y,\,\mathbf{1}\big]$ and $\beta=(a,b,c,d,e,f)^\top$,
the estimator is
\begin{equation}
\hat\beta_t \;=\; \arg\min_\beta \;\|W_t(A\beta - z)\|_2^2
\;=\; (A^\top W_t^2 A)^{-1}A^\top W_t^2 z,
\end{equation}
where $W_t=\mathrm{diag}(\,w_1,\dots,w_{|\mathcal{W}_t|}\,)$ is optional exponential weighting
(\emph{recent} samples weighted more), with $w_i\propto e^{-\alpha (t-s_i)}$.
Given $\hat a_t,\hat b_t,\hat c_t$, the \emph{local Gaussian curvature} at time $t$ is (Monge gauge approximation),\cite{CazalsPouget2005},
\begin{equation}
K_t \;\approx\; \frac{4\hat a_t \hat c_t - \hat b_t^2}{\big(1 + 4\hat a_t^2 + \hat b_t^2 + 4\hat c_t^2\big)^2}.
\label{eq:Kformula}
\end{equation}
Operationally, we pre-smooth $(x,y,z)$ (short moving average) and compute $K_t$ on an expanding or rolling window with a minimal sample size $m_0$ (see Appendix for defaults).

\begin{figure}[H]
\centering

\begin{minipage}{0.48\linewidth}\centering
  \includegraphics[width=\linewidth]{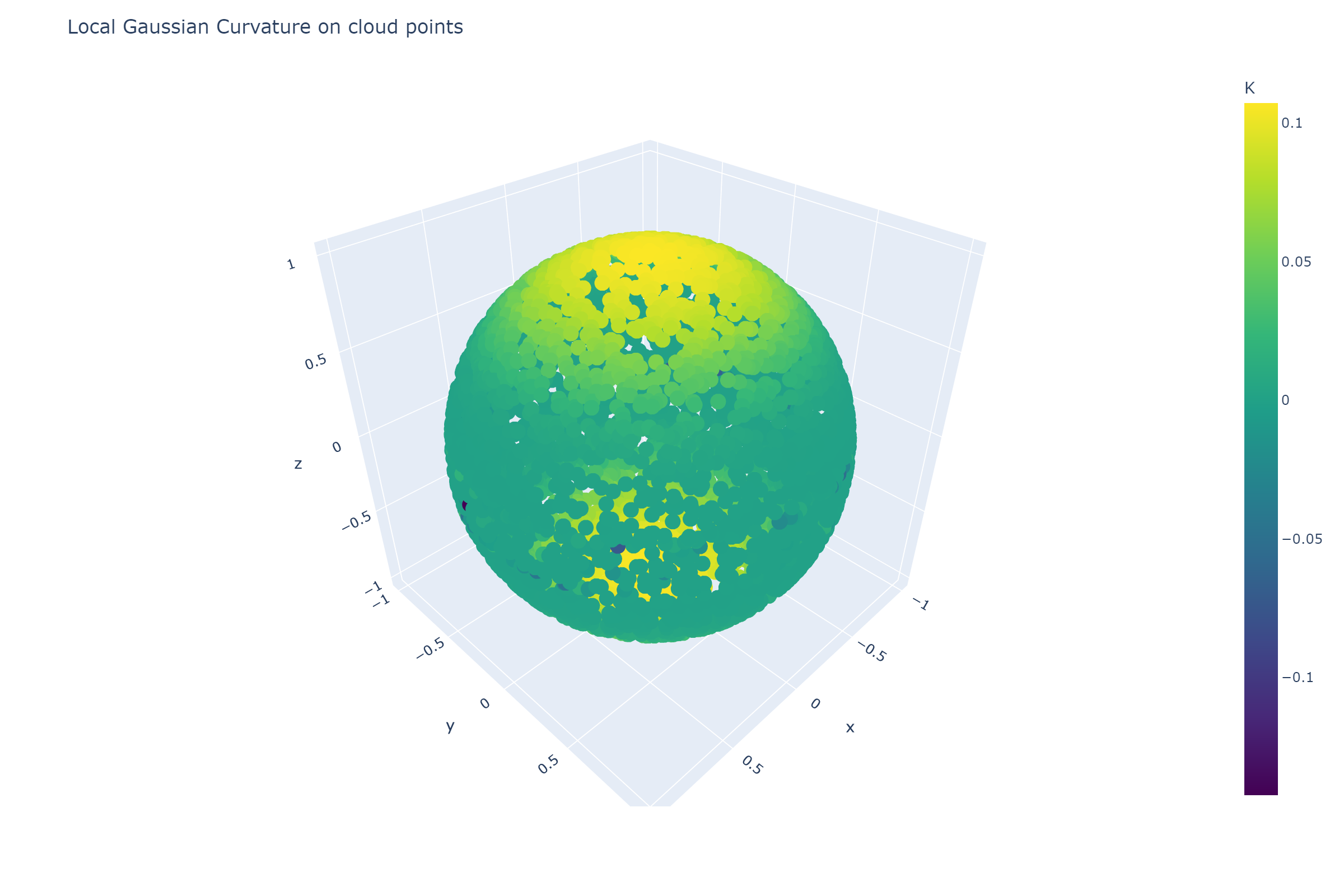}\\[-2pt]
  \small (a) Sphere $S^2$ (uniform sampling)
\end{minipage}\hfill
\begin{minipage}{0.48\linewidth}\centering
  \includegraphics[width=\linewidth]{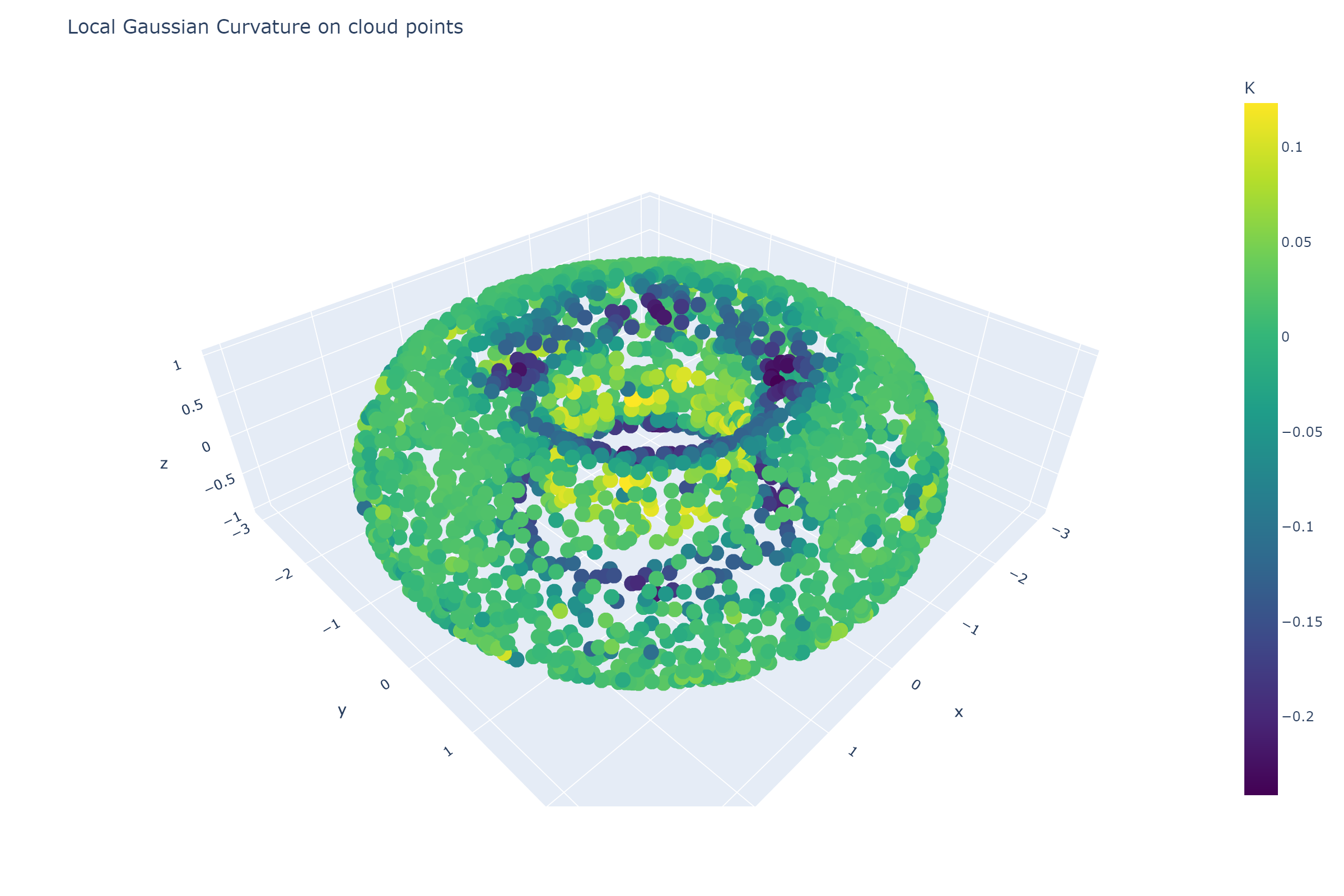}\\[-2pt]
  \small (b) Torus $T^2$ (uniform sampling)
\end{minipage}

\vspace{0.6em}

\begin{minipage}{0.48\linewidth}\centering
  \includegraphics[width=\linewidth]{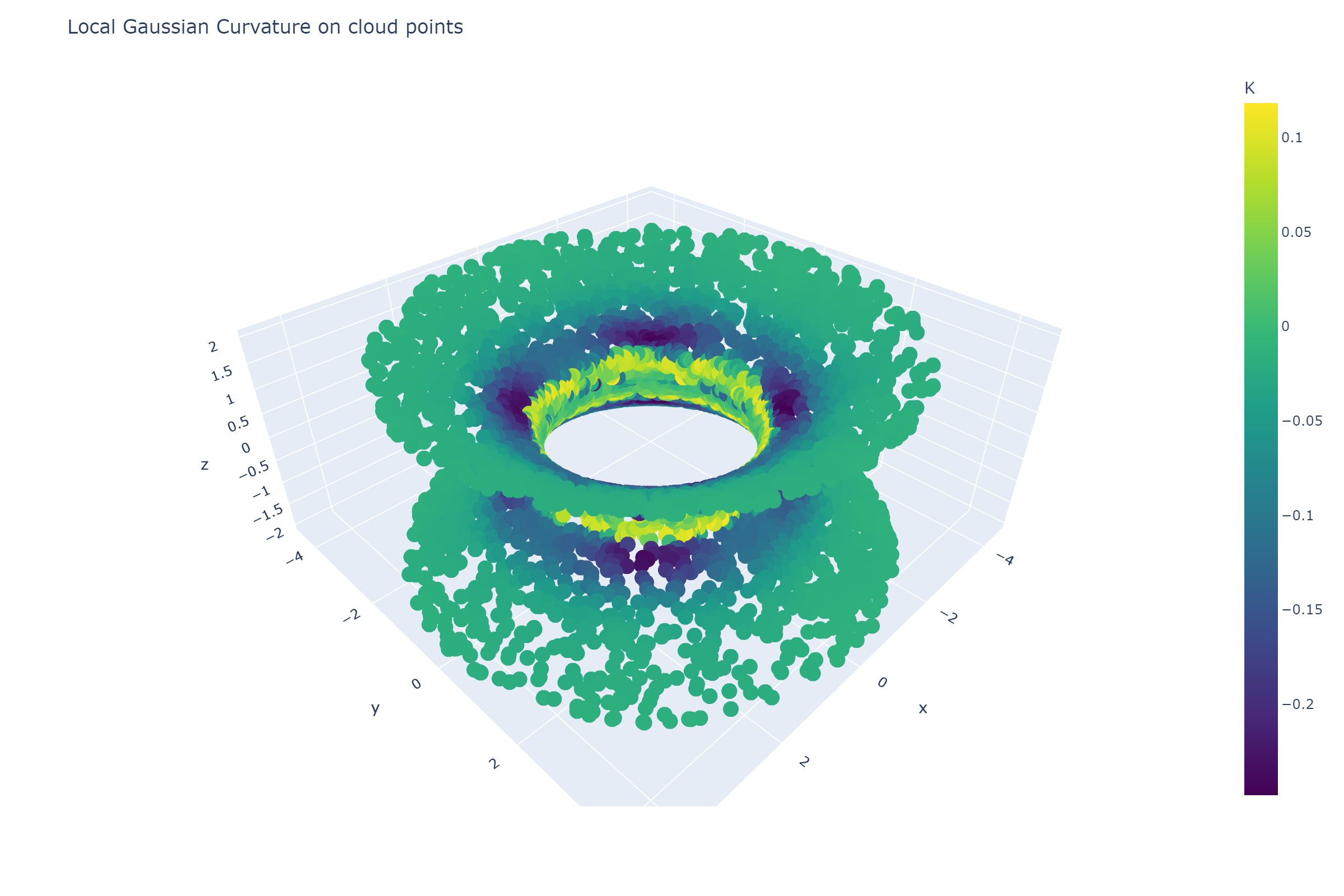}\\[-2pt]
  \small (c) Hyperboloid $H^2$ (uniform sampling)
\end{minipage}\hfill
\begin{minipage}{0.48\linewidth}\centering
  \includegraphics[width=\linewidth]{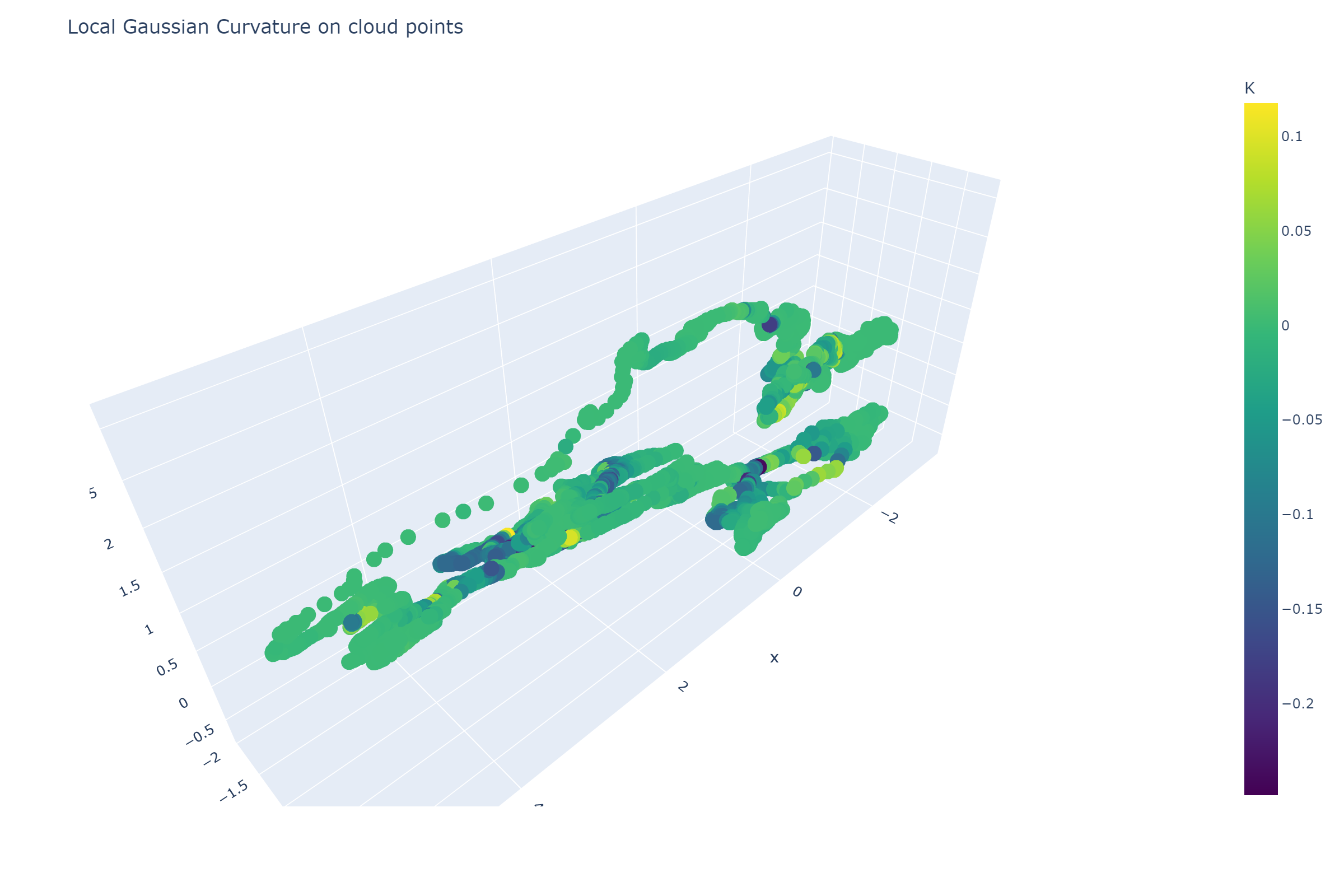}\\[-2pt]
  \small (d) Finance path (expanding-window PCA 3D)
\end{minipage}

\caption{Local Gaussian curvature estimates $K$ across benchmark shapes and the real-data embedded path. The benchmarks provide sign/scale references; the finance panel shows intermittent, regime-like curvature bursts.}
\label{fig:curv_multifigure}
\end{figure}

% --- One-liner (kept compact) ---
Towards validating that the method approximates satisfactorily the underlying Gaussian curvature, Fig.~\ref{fig:curv_multifigure} recovers the expected signatures on benchmarks—-$K>0$ on $S^2$, mixed $K$ on $T^2$, and $K<0$ on $H^2$—and yields plausible, intermittent curvature on the finance path; “uniformly sampled’’ means draws from each surface’s Riemannian (area) measure (density $\propto dA$).\cite{doCarmoRiemannian}

\medskip
\noindent\begin{minipage}{\linewidth}\small
\[
\begin{aligned}
\textbf{Sphere } S^2(R):\quad
& x(\theta,\varphi)=\big(R\cos\theta\,\sin\varphi,\; R\sin\theta\,\sin\varphi,\; R\cos\varphi\big),\\[-2pt]
& (\theta,\varphi)\in[0,2\pi)\times[0,\pi],\qquad
dA=R^2\sin\varphi\, d\theta\, d\varphi.
\end{aligned}
\]
\[
\begin{aligned}
\textbf{Torus } T^2(R,r):\quad
& x(\theta,\phi)=\big((R+r\cos\phi)\cos\theta,\; (R+r\cos\phi)\sin\theta,\; r\sin\phi\big),\\[-2pt]
& (\theta,\phi)\in[0,2\pi)^2,\qquad
dA=r\,(R+r\cos\phi)\, d\theta\, d\phi.
\end{aligned}
\]
\[
\begin{aligned}
\textbf{Hyperboloid}:\quad
& x(u,v)=\big(A\cosh u\cos v,\; A\cosh u\sin v,\; c\sinh u\big),\\[-2pt]
& v\in[0,2\pi),\; u\in[u_{\min},u_{\max}],\quad
dA= A\cosh u\,\sqrt{A^2\sinh^2 u + c^2\cosh^2 u}\, du\, dv.
\end{aligned}
\]
\end{minipage}

% --- Ultra-compact sampling recipes (inline; layout-safe) ---
\noindent\begin{minipage}{\linewidth}\small
\textit{Practical uniform sampling:}
(i) $S^2$: sample $U,V\!\sim\!\mathrm{Unif}(0,1)$, set $\theta=2\pi U$, $\cos\varphi=1-2V$. 
(ii) $T^2$: sample $\theta\!\sim\!\mathrm{Unif}[0,2\pi)$ and $\phi$ by rejection with target $\propto R+r\cos\phi$. 
(iii) Hyperboloid: sample $v\!\sim\!\mathrm{Unif}[0,2\pi)$ and $u$ on $[u_{\min},u_{\max}]$ with target proportional to the $u$–marginal of $dA$.
\end{minipage}

% --- drop-in multi-figure (no subfigure package needed) ---

\paragraph{Geometry decision from $K_t$.}
Since $K>0$ on spheres, $K<0$ on hyperbolic patches, and $K\approx 0$ in flat regions, we use thresholds $0<\kappa_+\ll 1$, $0<\kappa_-\ll 1$ to define
\[\text{\bf Sphere-like if, } K_t\ge \kappa_+,\quad \]
\[\text{\bf Hyperbolic-like if, } K_t\le -\kappa_-,\quad \]
\[\text{\bf Flat (Euclidean-like) if, }  |K_t|<\min(\kappa_+,\kappa_-). \]

However, tori are \emph{mixed-curvature} surfaces (both signs occur), so we complement $K_t$ with a topological test.

\paragraph{Topological validation via persistent homology (Torus test).}
We form a \emph{Takens embedding}, \cite{Takens1981}, over the window (\cite{Takens1981}) $\mathcal{W}_t$:
\begin{equation}
\mathcal{T}_t \;=\; \big[\,X_{s},\,X_{s-\tau},\,\dots,\,X_{s-(m-1)\tau}\,\big]\in\mathbb{R}^{3m},
\end{equation}
with delay $\tau$ and embedding dimension $m$. We compute Vietoris Rips persistent homology of $\mathcal{T}_t$ and count persistent 1-cycles (Betti $1$ features),\cite{EdelsbrunnerHarer2010,Ghrist2008}. A torus satisfies $\beta_1=2$ (and $\beta_2=1$),\cite{EdelsbrunnerHarer2010}, so we flag
\[
\text{\bf Torus-like if }\ \#\{\text{H}_1\text{ lifetimes}>\epsilon\}\ \ge 2\] 
\[\quad(\text{optionally: and }\#\{\text{H}_2\text{ lifetimes}>\epsilon\}\ge 1)\]

with persistence threshold $\epsilon$ calibrated to the scale of the point cloud,\cite{Gidea2018,Ismail2022,Arvanitis2024}. \emph{In the current context, we implement the basic $\beta_1$–based detector over sliding windows}.

\paragraph{Final geometry decision rule.}
For each window:

\[\text{If Torus-like: geometry }=T^2;\]
\[\text{else if } K_t\ge \kappa_+: S^2;\]
\[\text{else if } K_t\le -\kappa_-: H^2;\] 
\[\text{else: Euclidean.}\] 

\medskip
\noindent\textbf{Notes.}\;
(i) We use expanding windows for stability on real data and rolling windows in stress tests;
(ii) exponential weighting ($\alpha>0$) emphasizes recency in the quadratic fit;
(iii) the torus test can be run on the joint $(x,y,z)$ embedding or on the scalar curvature series $K_t$ (scalar Takens) with similar thresholds.

\subsection{Forecasting in manifold space and baseline comparison}
\label{sec:forecasting}

\paragraph{Manifold-aware forecasting}
\begin{enumerate}\itemsep 0.15em
\item \emph{Regime inference:} Infer geometry using local gaussian curvature information - section \ref{sec:curv_infer}
\item \emph{Tangent velocities:} Compute $v_t = P(X_{t-1})\,(X_t-X_{t-1})$ to approximate intrinsic velocity.
\item \emph{Tangent PCA:} Project $v_t$ on top-$d$ principal axes to obtain coefficients $c_t\in\mathbb{R}^d$.\cite{Jolliffe2002}
\item \emph{Time-series models:} Fit VAR($p$), to $\{c_t\}$ and forecast $\hat{c}_{t+1:t+h}$. We use rolling window VAR($p=25$) in our application.
\item \emph{Lift back:} Reconstruct $\hat{v}_{t+k}$ from $\hat{c}_{t+k}$, update $\hat{X}_{t+k}$ on $M$ via $\exp_{\hat{X}_{t+k-1}}(\hat{v}_{t+k})$ (or ambient step with projection), yielding path forecasts.
\end{enumerate}

\paragraph{Baseline (Native-space) forecasting — explicit comparator.}
We \emph{explicitly compare} against a geometry-agnostic baseline that applies the same predictors \emph{directly in the input space}: fit VAR to the raw $\mathbb{R}^3$ series $\{X_t\}$ (or its first differences) without manifold embeddings or tangent/PCA steps, and produce $\hat{X}^{\text{native}}_{t+1:t+h}$. All training windows, horizons, and hyperparameters are matched to ensure a fair comparison. \cite{LopezDePrado2018AFML,Jansen2023,Jolliffe2002}

\begin{figure}[H]
\centering
\resizebox{\linewidth}{!}{%
\begin{tikzpicture}[>=Latex]
\tikzstyle{blk}=[draw,rounded corners,align=center,inner sep=4pt,
                 text width=.22\linewidth,minimum height=9mm]

% top row
\node[blk, fill=gray!10] (data) {Input time series\\ $X_{1:T}\in\mathbb{R}^3$};
\node[blk, fill=blue!7,  right=5mm of data] (infer) {Geometry inference\\ Curvature $K_t$ + PH torus flag};
\node[blk, fill=green!7, right=5mm of infer] (tangent) {Tangent velocity\\ $v_t=P(X_{t-1})(X_t-X_{t-1})$};
\node[blk, fill=green!7, right=5mm of tangent] (pca) {Tangent PCA\\ $c_t\in\mathbb{R}^d$};

% second row
\node[blk, fill=yellow!15, below=8mm of pca] (models) {Forecasters\\ VAR / GP / RF on $c_t$};
\node[blk, fill=green!7,   right=5mm of models] (lift) {Lift back\\ $\hat v_{t+1}\!\rightarrow\! \hat X_{t+1}$};
\node[blk, fill=orange!20, right=5mm of lift] (alloc) {Allocation\\ Inverse-vol + tilt\\ Curvature gating};

% arrows
\draw[->] (data) -- (infer);
\draw[->] (infer) -- node[above]{$M$} (tangent);
\draw[->] (tangent) -- (pca);
\draw[->] (pca) -- (models);
\draw[->] (models) -- (lift);
\draw[->] (lift) -- (alloc);

% baseline
\node[blk, fill=red!7, below=8mm of tangent, text width=.30\linewidth] (native)
  {Baseline (native space)\\ Inverse-vol and VAR/GP/RF directly on $X_t$\\ No curvature/PH};
\draw[->, dashed] (data) |- (native);
\end{tikzpicture}%
}
\caption{Flow chart of the manifold-aware pipeline -- geometry ($M$) via curvature $K_t$ and persistent homology -- with an explicit native-space baseline.}
\end{figure}
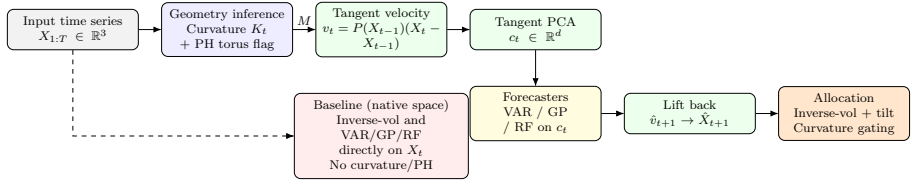

% ==== Figure 2 (overlap-safe): manifold log/exp + forecasting (titles inside boxes) ====
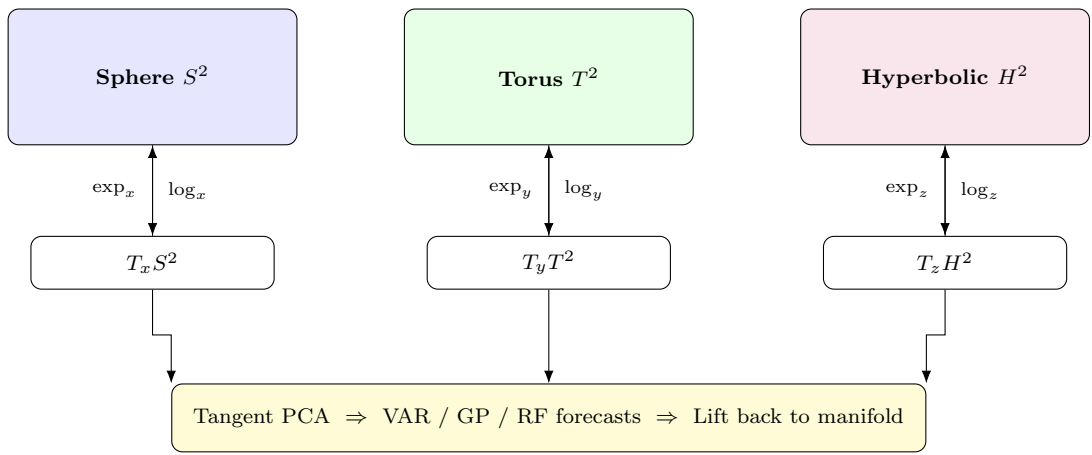
\begin{figure}[H]
\centering
\begin{tikzpicture}[>=Latex, font=\small]

% styles
\tikzset{
  geom/.style  ={draw, rounded corners, minimum height=18mm, minimum width=38mm,
                 align=center, fill=blue!10, fill opacity=1, inner sep=4pt},
  gtorus/.style={geom, fill=green!10},
  ghyp/.style  ={geom, fill=purple!10},
  plane/.style ={draw, rounded corners, fill=white, minimum height=7mm, minimum width=32mm, align=center},
  proc/.style  ={draw, rounded corners, fill=yellow!20, align=center,
                 minimum height=9mm, text width=9.5cm, inner sep=6pt}
}

% grid (matrix) to control spacing and avoid overlaps
\matrix (M) [matrix of nodes, row sep=12mm, column sep=14mm, nodes={anchor=center}]
{
  % Row 1: geometry boxes WITH labels inside
  \node[geom]  (sphere) {\textbf{Sphere $S^2$}}; &
  \node[gtorus](torus)  {\textbf{Torus $T^2$}}; &
  \node[ghyp]  (hyp)    {\textbf{Hyperbolic $H^2$}}; \\
  % Row 2: tangent planes
  \node[plane] (Ts) {$T_x S^2$}; &
  \node[plane] (Tt) {$T_y T^2$}; &
  \node[plane] (Th) {$T_z H^2$}; \\
};

% log/exp arrows (short, direct; keep away from labels)
\draw[->] (sphere.south) -- node[right=1mm]{\scriptsize $\log_x$} (Ts.north);
\draw[->] (Ts.north)     -- node[left =1mm]{\scriptsize $\exp_x$} (sphere.south);

\draw[->] (torus.south)  -- node[right=1mm]{\scriptsize $\log_y$} (Tt.north);
\draw[->] (Tt.north)     -- node[left =1mm]{\scriptsize $\exp_y$} (torus.south);

\draw[->] (hyp.south)    -- node[right=1mm]{\scriptsize $\log_z$} (Th.north);
\draw[->] (Th.north)     -- node[left =1mm]{\scriptsize $\exp_z$} (hyp.south);

% midpoint between Ts and Th (named coordinate so 'of' works)
\coordinate (midT) at ($(Ts)!0.5!(Th)$);

% PCA/forecast box spans the width under the planes
\node[proc, below=16mm of midT] (pcaF)
  {Tangent PCA $\;\Rightarrow\;$ VAR / GP / RF forecasts $\;\Rightarrow\;$ Lift back to manifold};

% feed lines to the PCA/forecast box (slightly angled to avoid crowding)
\draw[->] (Ts.south) -- ++(0,-6mm) -| (pcaF.north west);
\draw[->] (Tt.south) -- (pcaF.north);
\draw[->] (Th.south) -- ++(0,-6mm) -| (pcaF.north east);

\end{tikzpicture}
\caption{Manifold embedding (log), tangent-space forecasting, and lifting (exp). Labels are placed \emph{inside} the geometry boxes.}
\end{figure}

\begin{figure}[H]
  \centering
  \includegraphics[width=\linewidth]{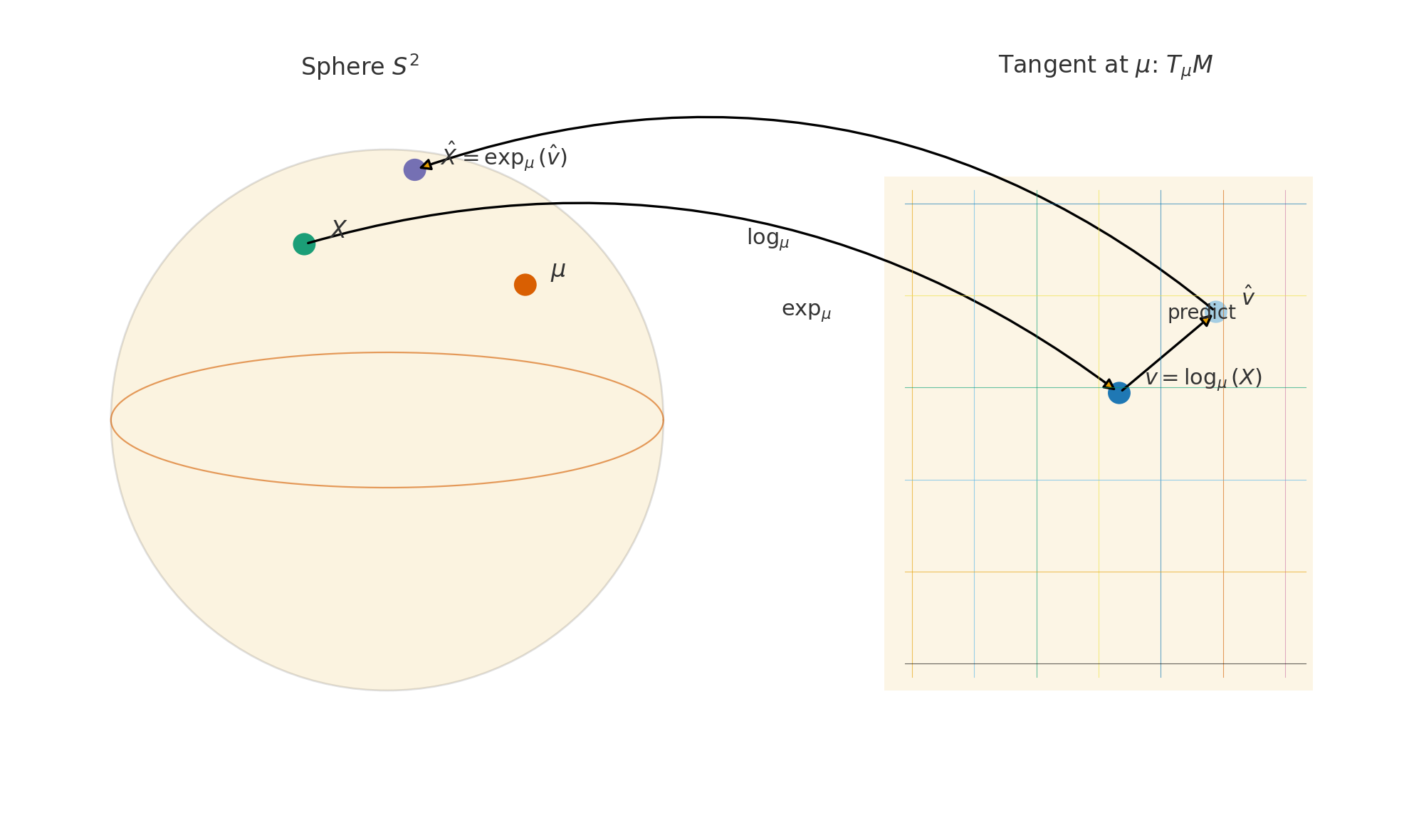}
  \caption{Sphere $S^2$: log map to tangent at $\mu$, prediction $\hat v$, and lift $\hat X=\exp_\mu(\hat v)$.}
\end{figure}

\begin{figure}[H]
  \centering
  \includegraphics[width=\linewidth]{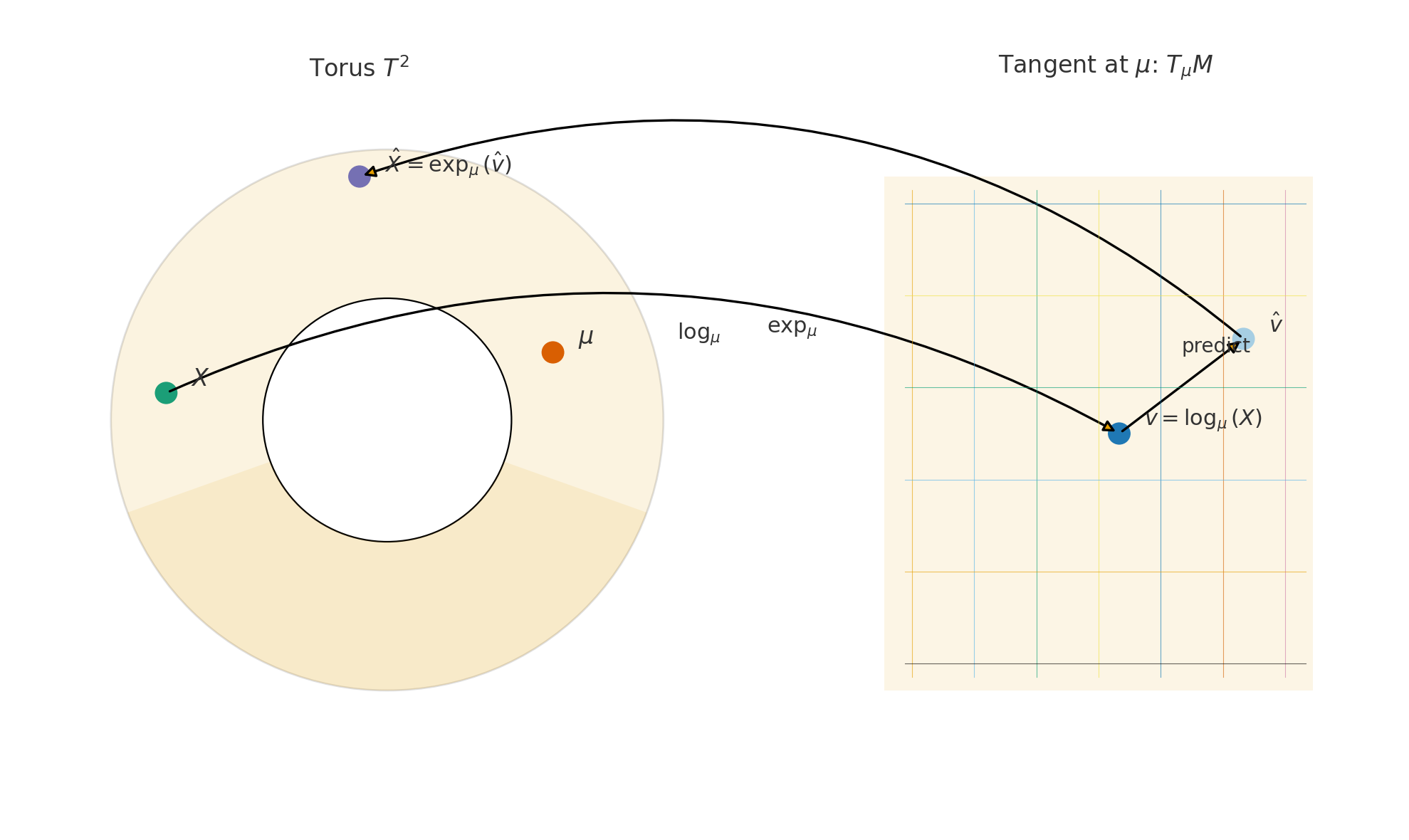}
  \caption{Torus $T^2$: log map, tangent-space prediction, and lifting back via $\exp_\mu$.}
\end{figure}

\begin{figure}[H]
  \centering
  \includegraphics[width=\linewidth]{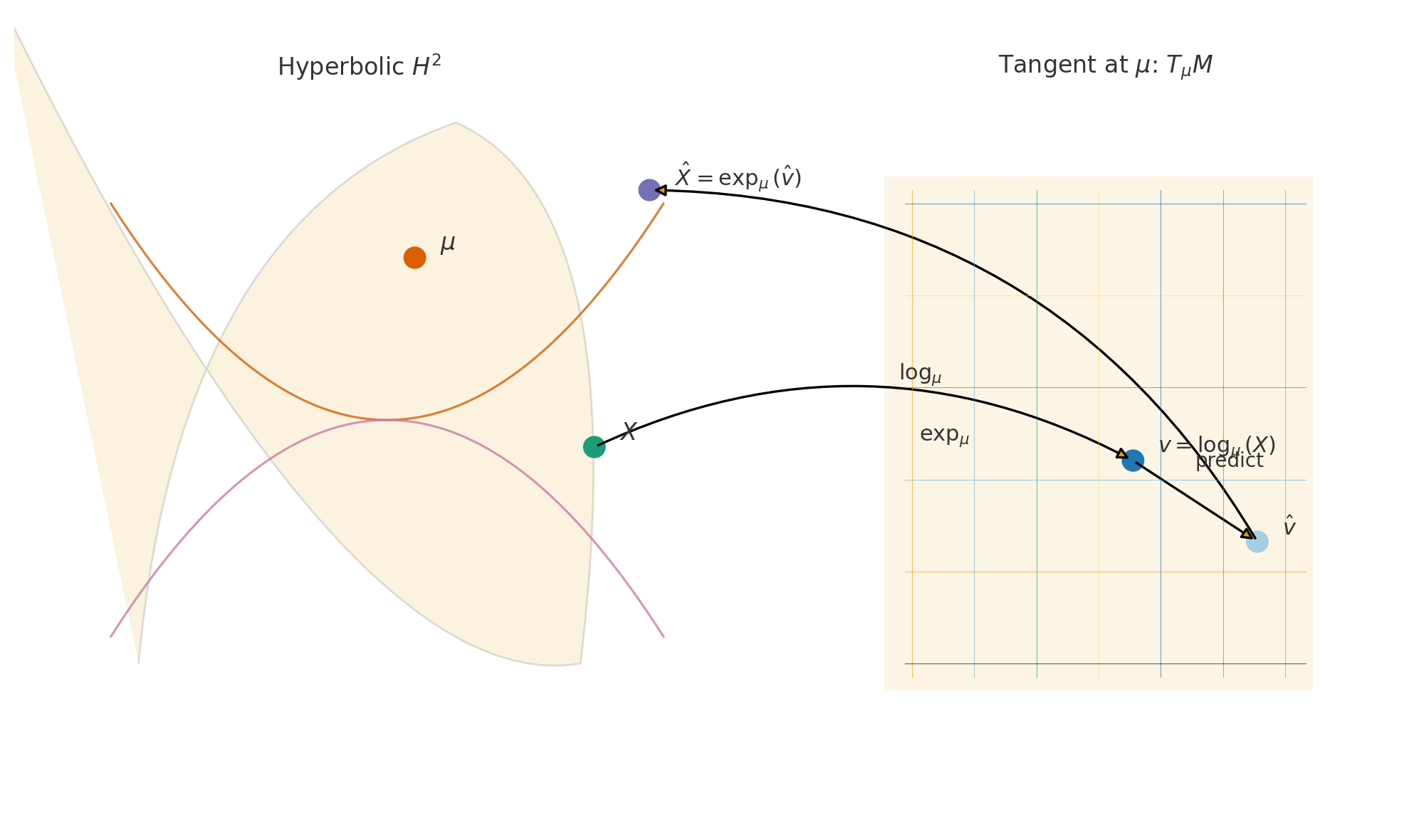}
  \caption{Hyperbolic $H^2$: log map, tangent-space prediction, and lifting via $\exp_\mu$.}
\end{figure}

\subsection{Euclidean Null Control: Correlated Brownian Motions}
\label{subsec:euclid_null_control}

To verify that our pipeline does \emph{not} confuse linear factor structure (e.g., PCA geometry) with \emph{intrinsic} manifold geometry, we run a Euclidean ``null'' experiment based on correlated Brownian motions. This control is designed to isolate what PCA can explain in a flat space and to demonstrate that our curvature/topology inference remains neutral (flat) when no curved geometry is present.

\paragraph{Construction (flat $\mathbb{R}^n$ model).}
Fix $n\ge 3$, horizon $T$, and an equicorrelation level $\rho\in[0,1)$.
Let
\begin{equation}
\Sigma_\rho \;=\; \rho\,\bm 1\bm 1^\top + (1-\rho)\,I_n
\qquad(\text{PSD, full-rank for } \rho<1),
\end{equation}
and let $L$ be a Cholesky factor of $\Sigma_\rho$.
Generate i.i.d.\ standard normal innovations $Z_t\sim\mathcal{N}(0,I_n)$ and set
\begin{equation}
\Delta W_t \;=\; L\,Z_t \;\sim\; \mathcal{N}(0,\Sigma_\rho),
\qquad
W_t \;=\; \sum_{s=1}^{t} \Delta W_s,\quad t=1,\dots,T.
\end{equation}
Thus $\{W_t\}$ is a \emph{multivariate Brownian motion in flat $\mathbb{R}^n$} with constant diffusion $\Sigma_\rho$ (no curvature, no manifold constraints).\cite{RevuzYor1999}

\paragraph{What this controls for.}
In the equicorrelation model, any structure is purely \emph{linear correlation}. If one performs PCA on the cross-section,
\begin{equation}
\lambda_1 \;=\; 1+(n-1)\rho, 
\qquad
\lambda_2=\cdots=\lambda_n \;=\; 1-\rho,
\end{equation}
revealing one dominant ``market'' factor and $n{-}1$ equal idiosyncratic directions. Projecting trajectories onto the top 2--3 PCs is a \emph{linear} rotation/scale; it does \emph{not} induce curvature.\cite{Jolliffe2002}
Hence, if our method were merely ``learning PCA's geometry,'' it would incorrectly report non-flat geometry here. The null control checks that it does not.

\paragraph{Embedding and diagnostics.}
To match the rest of our pipeline, we form a 3D time series $X_t\in\mathbb{R}^3$ either by:
(i) selecting three coordinates of $W_t$, or
(ii) projecting $W_t$ onto the first three PCs (purely linear compression).
On this 3D path we then compute:
\begin{enumerate}
\item \textbf{Local Gaussian curvature} $K_t$ via a weighted quadratic Monge patch fit
$z=ax^2+bxy+cy^2+dx+ey+f$ on rolling/expanding windows (Section~\ref{subsec:curv_stats_fin}).
\item \textbf{Topological torus test} via Takens embedding and persistent homology
(looking for two persistent $H_1$ cycles).
\end{enumerate}

\paragraph{Expected outcome under the null.}
Because the data live in an \emph{affine} (flat) subspace:
\begin{itemize}
\item $K_t \approx 0$ up to finite-sample noise. For any small thresholds $\kappa_+,\kappa_->0$,
\[
\mathbb{P}\big(|K_t|<\min(\kappa_+,\kappa_-)\big)\;\text{is high,}
\]
so the decision rule classifies as \emph{Euclidean/flat}.
\item The persistent homology \emph{does not} show two long-lived 1-cycles, hence no torus flag.
\end{itemize}

\paragraph{Forecasting and allocation implications.}
In a flat regime:
\begin{itemize}
\item Tangent-space PCA of increments and native-space modeling are \emph{effectively equivalent} (both are linear).
\item The curvature gating $\lambda_t$ satisfies $\lambda_t\simeq 1$ (no expansion/contraction).
\item Portfolio weights collapse to the baseline inverse-vol rule (plus any small return tilt), with no systematic advantage to manifold-aware steps.
\end{itemize}

This Euclidean null control demonstrates that our procedure \emph{does not} mistake PCA’s linear factor structure for intrinsic manifold geometry. Curvature/topology estimators remain flat when the data-generating process is flat, and any gains observed in the main experiments arise from genuine nonlinear geometric structure rather than artifacts of linear dimension reduction.

\subsection{Real-Finance Data Pipeline: Expanding PCA, Eigenportfolios, and Forecasting}
\label{subsec:real_fin_pipeline}

We apply our methodology to a broad multi-asset universe (equities, sectors, rates, credit, commodities, volatility indices; full tickers in Appendix~\ref{app:data_universe}). Raw daily prices (Yahoo Finance) ; log-returns are formed and basic long-only (LO) and risk-parity (RP) benchmark series are computed for reference. The LO and RP benchmarks (the latter built via inverse-volatility scaling) are saved alongside the panel of returns for later comparison.

\begin{figure}[H]
  \centering
  \includegraphics[width=\linewidth]{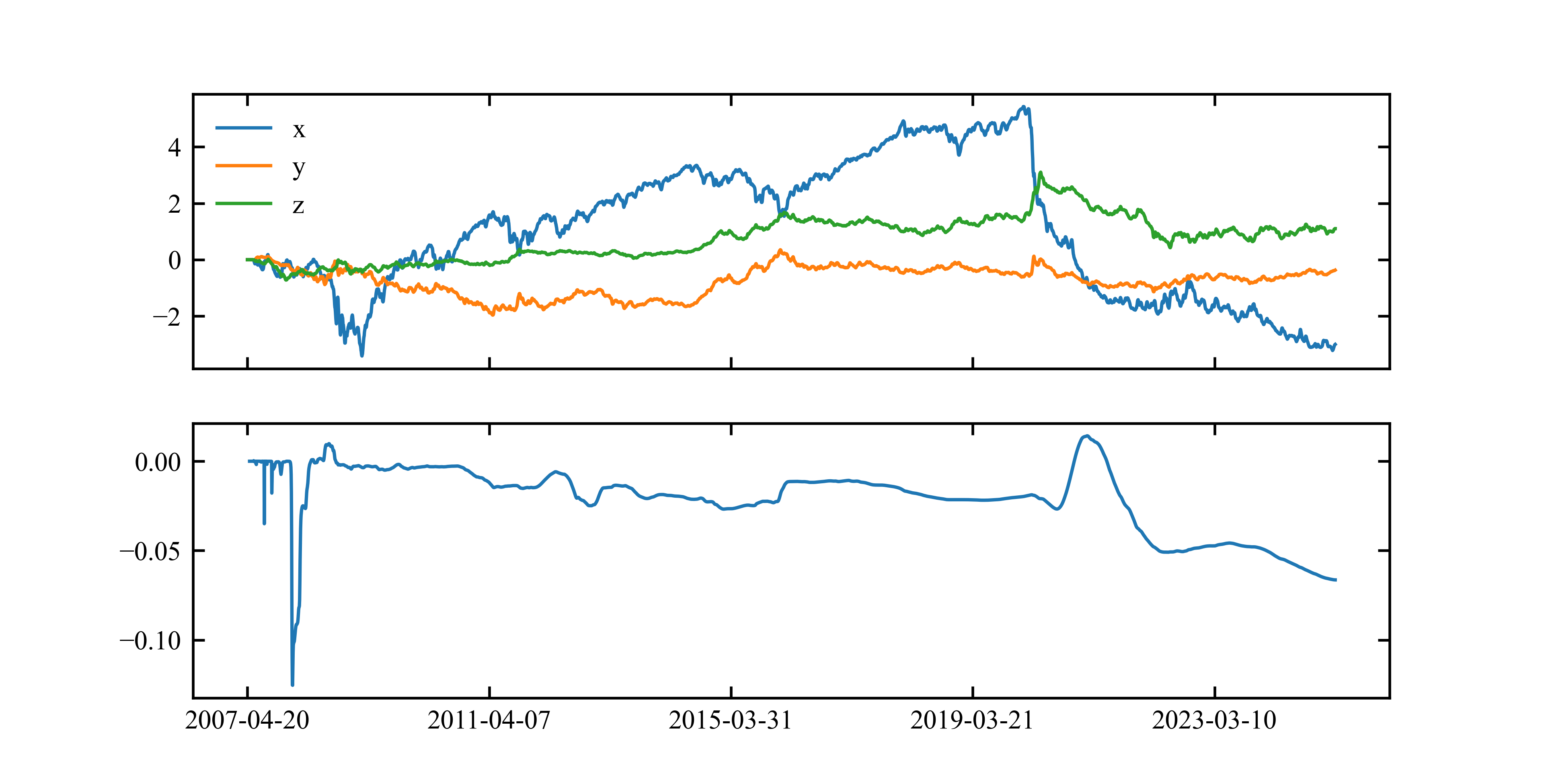}
  \caption{PCA projections evolution (Upper panel) and curvature estimation (Lower panel) for the real financial dataset. $x$ is PC1 (1st eigenportfolio), $y$ is the PC2 (2nd eigenportfolio) and $z$ the PC3 projection (3rd eigenportfolio), respectively}
  \label{fig:PCA_3D_Plot_and_Curvature}
\end{figure}

\paragraph{Expanding-window PCA and 3D embedding (eigenportfolios).}
Let $R_t\in\mathbb{R}^N$ denote the cross-section of asset returns at time $t$. For $t\ge t_0$, we standardize the history $\{R_1,\dots,R_t\}$ and compute PCA loadings $\{u_{1,t},u_{2,t},u_{3,t}\}$ and eigenvalues $\{\lambda_{1,t},\lambda_{2,t},\lambda_{3,t}\}$ on an \emph{expanding} window (\cite{Jolliffe2002,Avellaneda2022,LopezDePrado2020MLAM}. To avoid look-ahead, eigenportfolio $i$ at time $t$ uses the \emph{lagged} loadings:
\begin{equation}
p_{i,t} \;=\; u_{i,t-1}^{\top} R_t,\qquad i=1,2,3,
\label{eq:eigenport_rets}
\end{equation}
so $p_{i,t}$ is the (out-of-sample) return of the $i$th eigenportfolio. Stacking the three principal sleeves yields a \emph{3D embedded path} in cumulative form
\begin{equation}
X_t \;=\; \Big(\,\sum_{s\le t}p_{1,s},\ \sum_{s\le t}p_{2,s},\ \sum_{s\le t}p_{3,s}\,\Big)^\top \in\mathbb{R}^3,
\label{eq:3d_path}
\end{equation}
which serves as the input trajectory for our geometry-aware predictor. We display the expanding window curvature estimation for the PCA embedded financial dataset in Figure \ref{fig:PCA_3D_Plot_and_Curvature}

\paragraph{Geometry signal on finance: local curvature on the PCA path.}
On the 3D path $X_t$ we estimate the \emph{local Gaussian curvature} $K_t$ by fitting quadratic Monge patches on rolling/expanding neighborhoods (Section~\ref{subsec:curv_stats_fin}). This series feeds the allocation rules and the curvature-aware benchmarks reported later.

\paragraph{Forecasting on the embedded path.}
Given $X_{1:t}$, we forecast the next embedded point $\widehat{X}_{t+1}$ either (i) \emph{natively in the 3D Euclidean path} or (ii) \emph{geometry-aware} by choosing a geometry $M\in\{\mathbb{R}^2,S^2,T^2,H^2\}$, mapping to the tangent space via $\log_\mu$, forecasting principal tangent coefficients (VAR / GP / RF), and lifting back via $\exp_\mu$ (Section \ref{subsec:curv_stats_fin}). 

\paragraph{Translating forecasts to trading signals and PnL.\cite{Maillard2010,Roncalli2013}}
Let $\Delta X_{t+1} = X_{t+1}-X_t$ be realized 3D increments and $\widehat{\Delta X}_{t+1}=\widehat{X}_{t+1}-X_t$. We form directional, volatility-scaled signals on each coordinate $k\in\{x,y,z\}$ as
\begin{equation}
s_{k,t+1}\;=\;\frac{\operatorname{sign}(\widehat{\Delta X}_{k,t+1})}{\widehat{\sigma}_{k,t}}
\label{eq:signal_rule}
\end{equation}

\begin{equation}
% 500-day rolling sample standard deviation (Bessel-corrected), for t \ge 500
\bar{\Delta X}_{k,t}^{(500)}
= \frac{1}{500}\sum_{j=0}^{499} \Delta X_{k,t-j},
\qquad
\widehat{\sigma}_{k,t}^{(500)}
= \sqrt{\frac{1}{499}\sum_{j=0}^{499}\!\left(\Delta X_{k,t-j}-\bar{\Delta X}_{k,t}^{(500)}\right)^{2}}\,.
\label{eq:sample_roll_vol}
\end{equation}

and computes coordinate PnL by $\,\mathrm{pnl}_{k,t+1}=s_{k,t+1}\,\Delta X_{k,t+1}\,$. 
\\
\\
Annualized Sharpe ratios, $\mathrm{Sh}[]$, are reported as
\begin{equation}
\mathrm{Sh}[\mathrm{pnl}_k]\;=\;\sqrt{252}\,\frac{\overline{\mathrm{pnl}}_k}{\mathrm{stdev}(\mathrm{pnl}_k)}\,,
\qquad
\mathrm{pnl}_{\text{Tot},t}\;=\;\sum_{k\in\{x,y,z\}}\mathrm{pnl}_{k,t}.
\label{eq:sharpe_rule}
\end{equation}

\subsection{Eigenvalue-Weighted Sleeves from \emph{3D PCA Space} (Expanding SVD)}
\label{subsec:eigval_weighting_3d}

In the finance application we first build the \emph{embedded 3D PCA path} of eigenportfolios
(Section~\ref{subsec:real_fin_pipeline}): for each date $t$ we have
\[
X_t \;=\; \big(X_{1,t},\,X_{2,t},\,X_{3,t}\big)^\top \in \mathbb{R}^3,
\]
where the coordinates are the out-of-sample cumulative eigenportfolio sleeves (PC1, PC2, PC3). \emph{Then}, \textbf{within this 3D space}, we compute an \emph{expanding-window SVD/PCA} to obtain time-varying variance levels and use those \emph{3D-space eigenvalues} to weight the sleeves,\cite{Jolliffe2002}.

Let $\Delta X_s := X_s - X_{s-1}$ and fix an expanding window $\mathcal{W}_t=\{1,\dots,t\}$.
Define the $3\times 3$ sample covariance on $\mathcal{W}_t$,
\[
\Sigma^{(3D)}_t \;=\; \frac{1}{|\mathcal{W}_t|}\sum_{s\in\mathcal{W}_t}\big(\Delta X_s - \overline{\Delta X}_t\big)\big(\Delta X_s - \overline{\Delta X}_t\big)^\top,
\qquad
\overline{\Delta X}_t = \frac{1}{|\mathcal{W}_t|}\sum_{s\in\mathcal{W}_t}\Delta X_s.
\]
Compute the spectral decomposition (equivalently, SVD of the $3\times |\mathcal{W}_t|$ matrix of $\Delta X_s$)
\[
\Sigma^{(3D)}_t \;=\; Q_t\,\Lambda_t\,Q_t^\top,\qquad
\Lambda_t=\mathrm{diag}\!\big(\lambda_1(t),\lambda_2(t),\lambda_3(t)\big),\;\; \lambda_1(t)\!\ge\!\lambda_2(t)\!\ge\!\lambda_3(t)\!\ge\!0.
\]
These $\lambda_i(t)$ are the \emph{expanding-window eigenvalues in the 3D PCA space} 

\paragraph{Eigenvalue-driven sleeve weights in 3D space.}
We map the variance levels into normalized sleeve weights
\begin{equation}
C_{i,t} \;=\; \frac{\lambda_i(t)}{\lambda_1(t)+\lambda_2(t)+\lambda_3(t)} \,,\qquad i=1,2,3,
\label{eq:eig_weights_3d}
\end{equation}
so that directions with larger expanding-window energy in the 3D PCA space receive higher allocation.

\paragraph{Forecast integration and portfolio return.}
Let $\widehat{X}_{t+1}$ be the predicted point from either (i) the \emph{native-space} forecaster, or
(ii) the \emph{geometry-aware} (log–forecast–exp) forecaster; define
$\widehat{\Delta X}_{i,t+1}=\widehat{X}_{i,t+1}-X_{i,t}$ and the directional signal
$s_{i,t+1}=\mathrm{sign}(\widehat{\Delta X}_{i,t+1})$. The eigenvalue-weighted eigenportfolio return is
\begin{equation}
r^{(\mathrm{eig},\,3D)}_{t+1}
\;=\;\sum_{i=1}^3 C_{i,t}\; s_{i,t+1}\; p_{i,t+1},
\qquad
p_{i,t+1} \text{ the out-of-sample return of sleeve } i.
\label{eq:eig_pnl_3d}
\end{equation}

All evaluation metrics (MAE/RMSE/Sign, Sharpe, cumulative PnL) are computed identically across (i) and (ii) to ensure a fair comparison; results are reported in Section~\ref{sec:results}.\cite{SimonianLopezdePradoFabozzi2024}

\paragraph{Design intent.}
This makes the \emph{allocation} responsive to the \emph{expanding-window variance structure \emph{of the 3D PCA space itself}} (via $\lambda_i(t)$), while the \emph{forecasting layer} tests whether geometry-aware predictions improve the \emph{directional timing} $s_{i,t+1}$ relative to a Euclidean baseline.

\paragraph{Curvature-aware aggregation and geometry-weighted benchmarks.}
Beyond the pure RP eigenportfolios framework, we also report:  
(i) a \emph{curvature-gated} aggregation that buckets timestamps by $K_t$ (negative/near-zero/positive) and averages the appropriate geometry-run PnLs (torus / Euclidean / all geometries) and
(ii) expanding-window geometry weighting by ex-post Sharpe/returns as a sanity check. These appear in the merged report alongside LO and RP asset benchmarks. 

\medskip
\noindent\textbf{Design intent.}
This setup ensures the following:  
(i) the 3D embedding reflects \emph{time-varying} linear structure (eigenportfolios) while all \emph{nonlinear} effects are captured by curvature/topology on the embedded path;  
(ii) forecasts are compared \emph{like-for-like} against a native-space baseline that applies the same predictors without manifold steps; and  
(iii) portfolio construction is neutral (inverse-vol) and modular curvature gating is reported separately so its incremental value can be isolated.

\section{Results}
\label{sec:results}

\subsection{'Forecast to Trading' Evaluation Design}
\label{subsec:eval_design}
We assess the methodology on (i) \emph{simulated} regimes and (ii) \emph{real finance} data (Section~\ref{subsec:real_fin_pipeline}). Because trading payoff is highly sensitive to \emph{direction}, we report both statistical errors (MAE/RMSE) and trading metrics (Sharpe/Sortino/Calmar, hit-rate). All comparisons are \emph{like-for-like} between:
\begin{itemize}
  \item \textbf{Native-space} forecasts in the 3D PCA embedding, and
  \item \textbf{Geometry-aware} forecasts (log–-forecast-–exp) on $M\in\{\mathbb{R}^2,S^2,T^2,H^2\}$.
\end{itemize}
Inputs, windows, and forecaster class are held fixed across arms.

\subsection{Curvature statistics and regime assignment (finance path)}
\label{subsec:curv_stats_fin}

Using the expanding-window estimator on the 3D PCA path, we analyze the series $\{K_t\}$ (4491 non-missing points, from 2007 to 2025). Basic distributional facts:
\[\text{mean}(K)=-0.0207,\quad \text{sd}(K)=0.0186,\quad\]
\[\min K=-0.125,\quad \mathrm{Q}_{25}=-0.0247,\ \mathrm{Q}_{50}=-0.0186,\ \]
\[\mathrm{Q}_{75}=-0.0088,\ \max K=0.0142.\]

With curvature thresholds $(\kappa_+,\kappa_-)=(+0.01,-0.01)$, the time share by regime is:
\[
\underbrace{\mathbb{P}(K_t\le \kappa_-)}_{\text{hyperbolic-like}}=73.9\%,\qquad
\underbrace{\mathbb{P}(|K_t|\le 0.01)}_{\text{near-flat}}=24.4\%,\qquad
\underbrace{\mathbb{P}(K_t\ge \kappa_+)}_{\text{spherical-like}}=1.7\%.
\]
At looser thresholds (e.g., $\tau=0.005$) the near-flat share rises to $18.4\%$ and the positive share remains small ($3.3\%$), while $78.3\%$ remains negative. Serial dependence is strong (ACF(1)$=0.997$), indicating \emph{persistent} curvature regimes rather than high-frequency noise.

\paragraph{Time segmentation (annual shares, $\tau=0.01$).}
A compact, label-based classification by year assigns “hyperbolic-like” when $\mathbb{P}(K_t\le-0.01)>\!50\%$ and $\mathbb{P}(K_t\ge 0.01)<10\%$; “Euclidean/flat-like” when $\mathbb{P}(|K_t|\le 0.01)>\!60\%$ and both tails are small; and “torus-like (mixed)” when both tails are material ($>\!10\%$ each). The resulting picture is:

\begin{itemize}
\item \textbf{2007–-2010:} Euclidean/flat-like (near-flat share dominates; mean $K$ mild negative).
\item \textbf{2011–--2020:} Hyperbolic-like (negative curvature dominates; virtually no positive tail).
\item \textbf{2021:} \emph{Torus-like (mixed)} — $\mathbb{P}(K_t\le-0.01)\approx 48\%$, $\mathbb{P}(|K_t|\le0.01)\approx 25\%$, $\mathbb{P}(K_t\ge 0.01)\approx 26\%$.
\item \textbf{2022–-2025:} Strongly hyperbolic-like (negative share $\approx 100\%$, increasingly negative mean $K$).
\end{itemize}

\paragraph{Interpretation.}
At face value, the \emph{sign} distribution is skewed negative (hyperbolic-like), but two features align with a \emph{toroidal scaffold} dominating the overall dynamics:
(i) the presence of both negative and positive curvature episodes (albeit asymmetric), and
(ii) extended stretches of near-flat curvature between negative bursts.
On a standard torus $T^2(R,r)$ the Gaussian curvature varies with the minor angle $\phi$,
\[
K(\phi)=\frac{\cos\phi}{r\,[R+r\cos\phi]},
\]
so trajectories that spend more time near the inner saddle ($\phi\!\approx\!\pi$) naturally produce a distribution with large negative mass, occasional small positive excursions (outer bulge, $\phi\!\approx\!0$), and plateaus near $K\!\approx\!0$ when the path lingers on transition bands. This is exactly what we observe: long negative runs (median run length $\approx 14$ time points; max $\sim 3000$), short positive runs (median $\approx 3$), and nontrivial flat intervals. 

\medskip
\noindent\textbf{Assignment.} Aggregating across 2007 to 2025, the curvature statistics point to \emph{hyperbolic-biased toroidal dynamics}: topologically consistent with one (or multiple) torii, but with the path predominantly visiting the saddle side (negative $K$), punctuated by mixed-curvature episodes (notably 2021) and near-flat interludes. This reconciles (a) the dominance of negative curvature in the estimator with (b) the empirical finding that toroidal embeddings often deliver the strongest predictive sleeves and portfolio lift when those mixed/periodic phases emerge.
% =========================
% Topological validation (REAL DATA) — structure + placeholders
% =========================
\subsection{Topological validation via persistent homology }
\label{subsec:ph_real}

We provide a topological check for \emph{toroidal} behavior in the real-data embedding by estimating the homology of a delay-embedded attractor (Takens’ embedding, \cite{Takens1981}) built from the 3D PCA path (and, where informative, from individual sleeves). The exhibit comprises: (i) a Takens attractor plot and (ii) persistent homology summaries (diagrams/barcodes) up to $H_2$. The torus test is \emph{heuristic but informative}: a $T^2$-like attractor typically shows one connected component ($\beta_0\!=\!1$), \emph{two} long-lived 1-cycles ($\beta_1\!\approx\!2$), and, in dense surface samples, a weak 2-class ($\beta_2\!\approx\!1$). Trajectory samples often recover the two $H_1$ classes cleanly, with $H_2$ less stable.

\paragraph{Takens embedding and filtration.}
Given a univariate coordinate $x_t$ (e.g., a PCA sleeve) or a multivariate series $X_t\in\mathbb{R}^3$, we construct a delay map
\[
\mathcal{T}_{m,\tau}(x_t)
= \big(x_t,\ x_{t-\tau},\ \dots,\ x_{t-(m-1)\tau}\big)\in\mathbb{R}^m,
\]
(or stack sleeves to get $\mathbb{R}^{3m}$), choose $(m,\tau)$ via standard criteria (mutual information / first ACF minimum; false nearest neighbors), standardize the cloud, and compute Vietoris–Rips persistence up to dimension 2. We summarize $H_k$ by lifetimes $\ell=b-d$ (birth $b$, death $d$), the number of long bars, and null-comparisons.

\begin{figure}[H]
\centering
\begin{minipage}{0.48\linewidth}\centering
  \includegraphics[width=\linewidth]{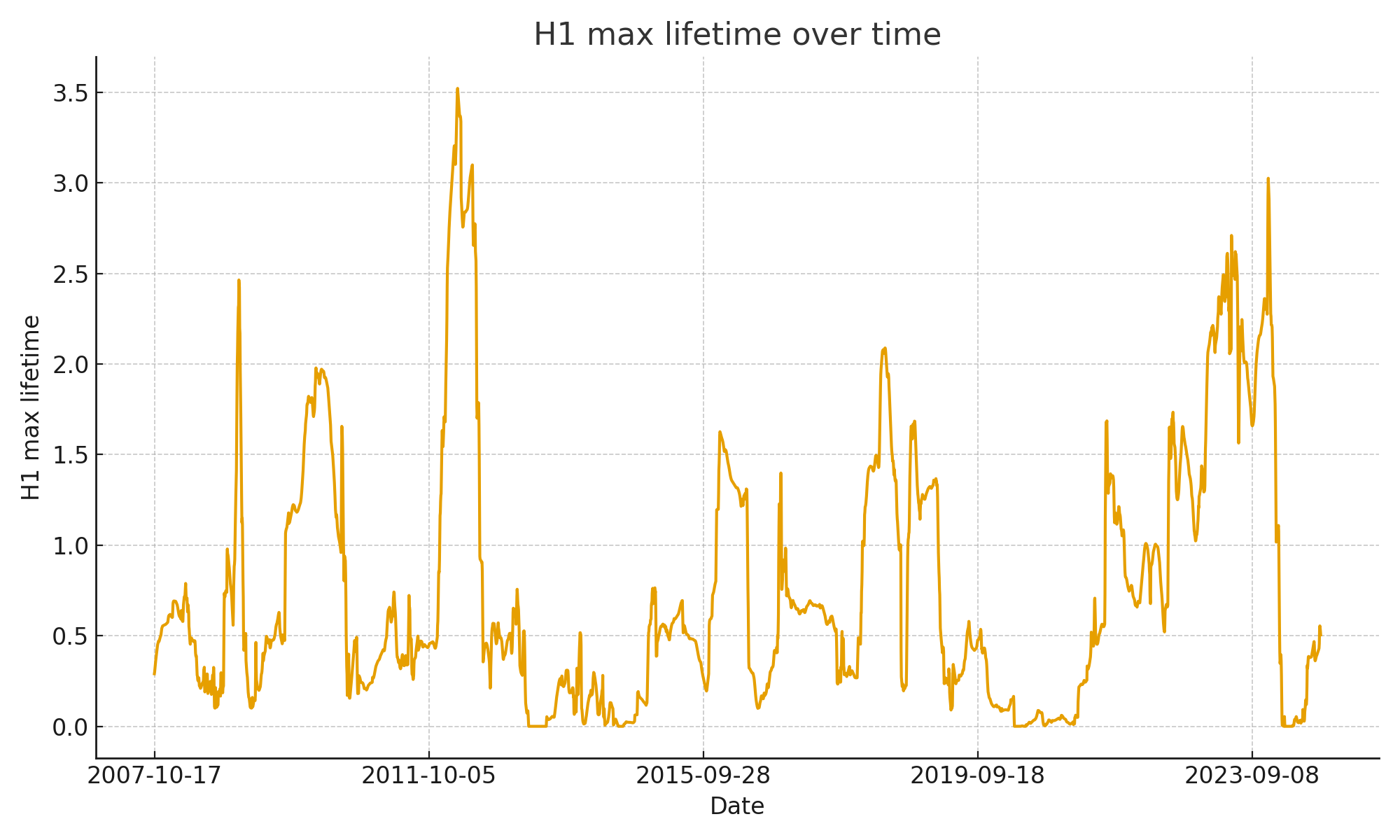}\\[-2pt]
  \small (a) H1 max lifetime (time series)
\end{minipage}\hfill
\begin{minipage}{0.48\linewidth}\centering
  \includegraphics[width=\linewidth]{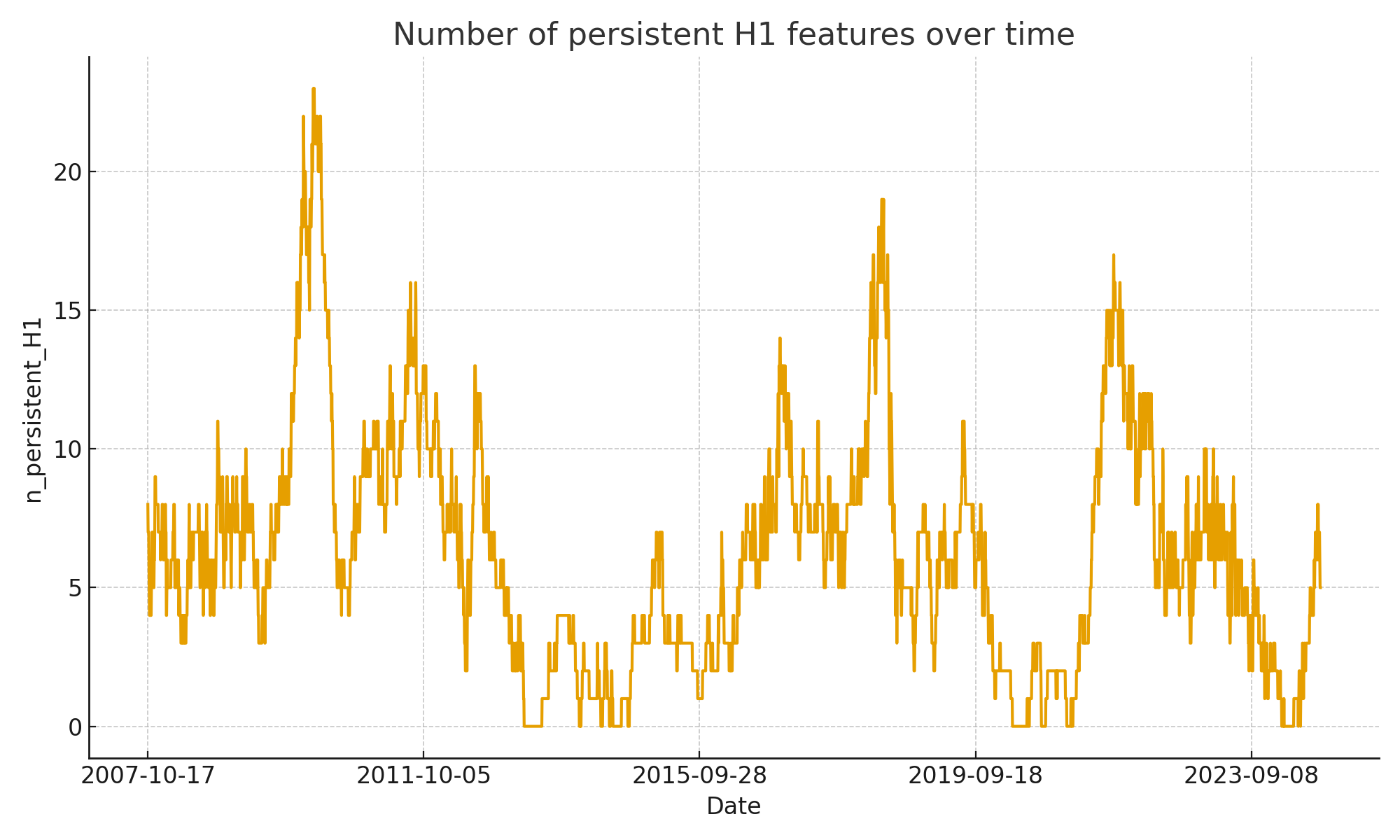}\\[-2pt]
  \small (b) \# persistent H1 loops (time series)
\end{minipage}

\vspace{0.6em}

\begin{minipage}{0.48\linewidth}\centering
  \includegraphics[width=\linewidth]{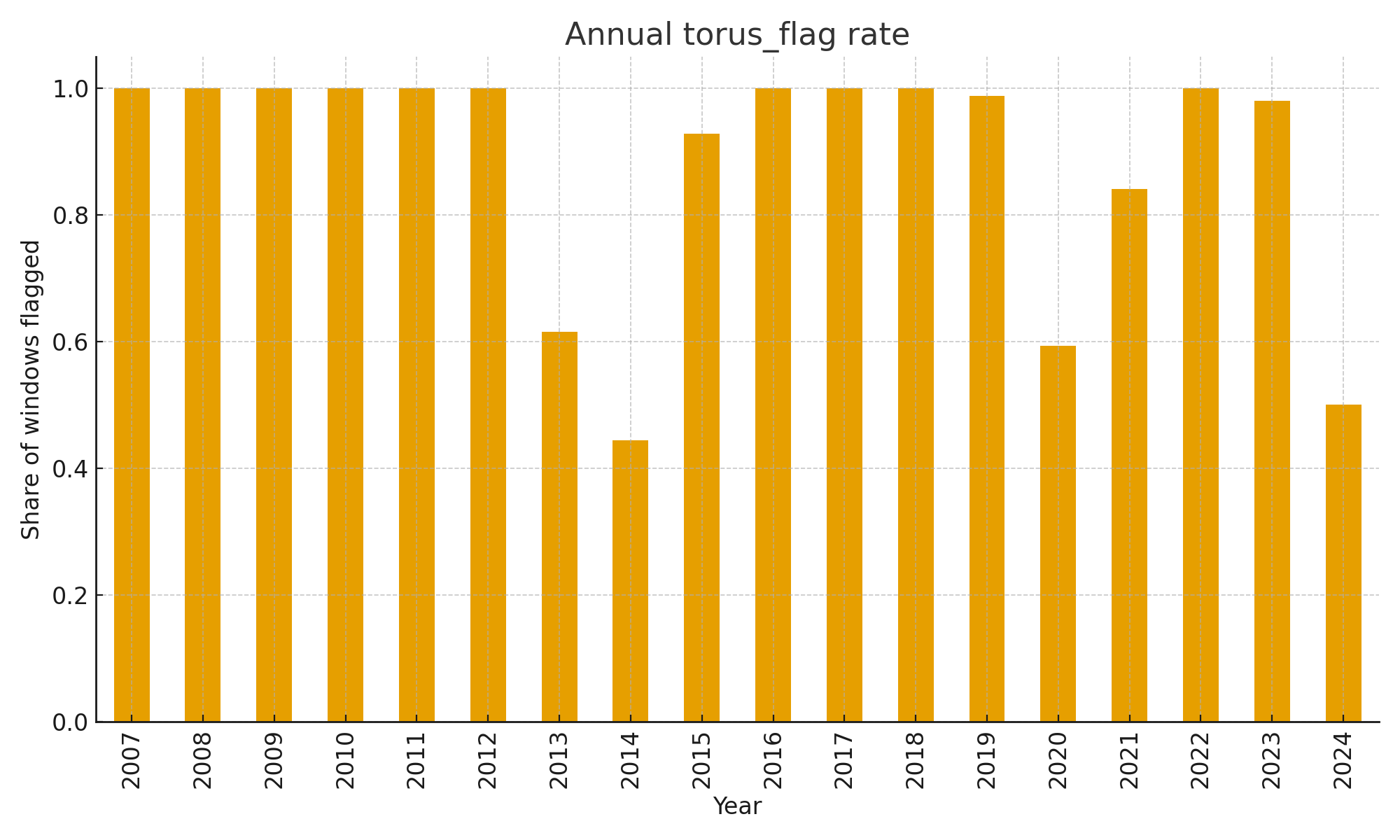}\\[-2pt]
  \small (c) Annual share of torus-flagged windows
\end{minipage}\hfill
\begin{minipage}{0.48\linewidth}\centering
  \includegraphics[width=\linewidth]{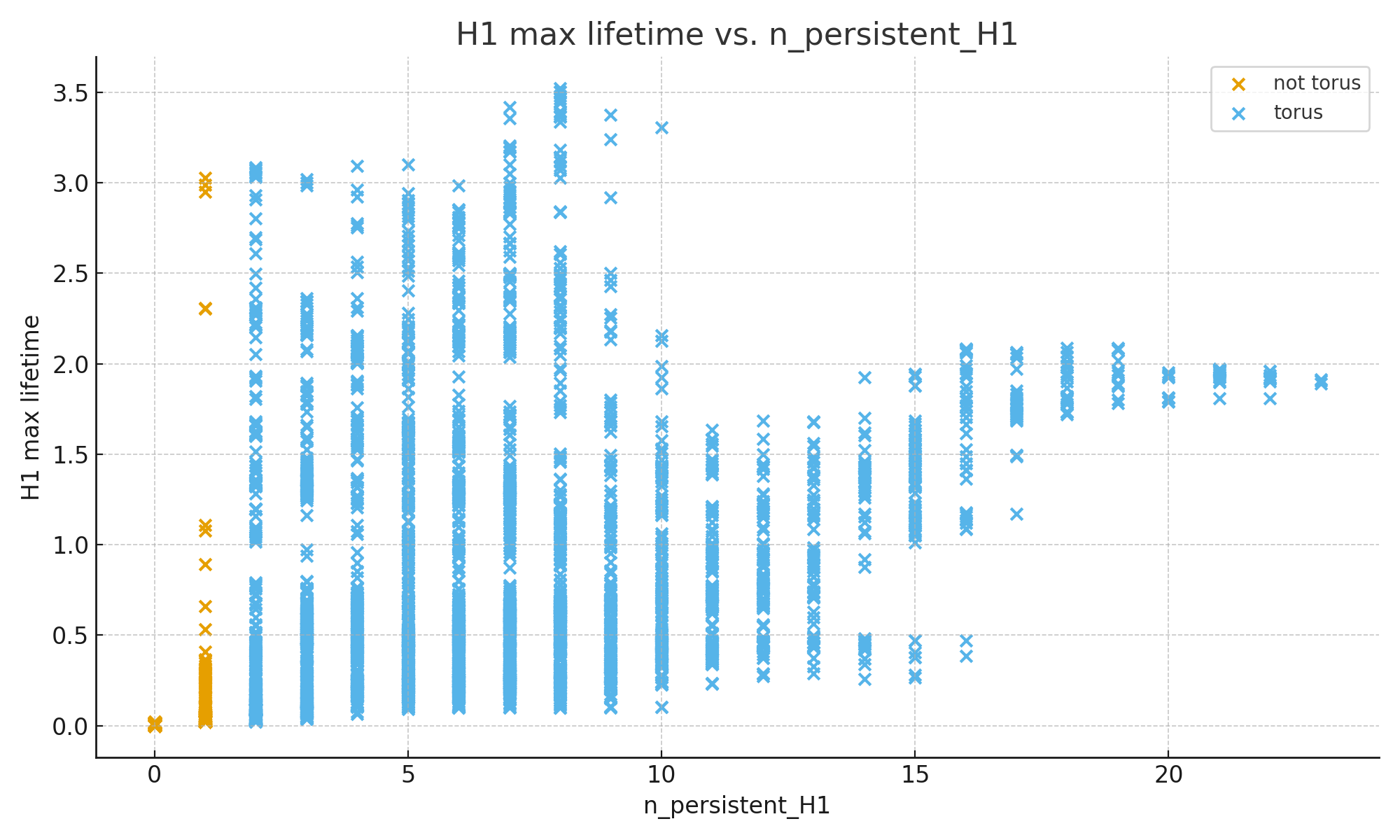}\\[-2pt]
  \small (d) H1 max lifetime vs \#H1 loops (flag split)
\end{minipage}

\caption{Persistent homology diagnostics on the real-data embedding. High \emph{H1} lifetimes (a) co-occur with elevated counts of persistent 1-cycles (b); year-by-year, torus-like intervals dominate (c). The moderate positive association in (d) indicates that when the two principal loops of a torus persist strongly, overall 1-dimensional topological activity is also elevated.}
\label{fig:ph_real_diag}
\end{figure}

\begin{figure}[H]
  \centering
  \includegraphics[width=\linewidth]{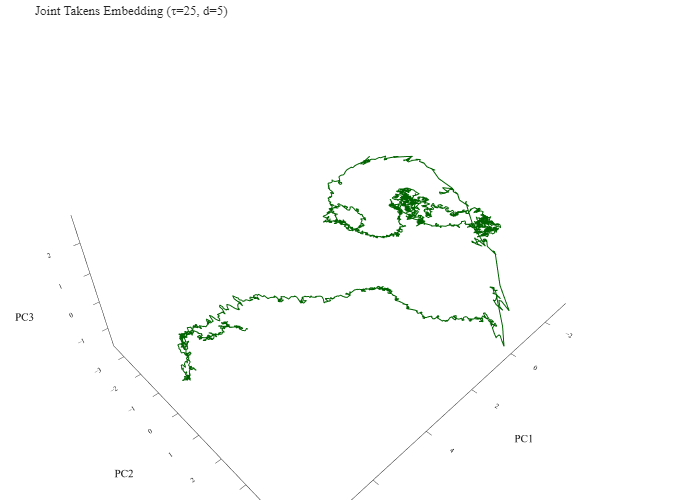}
  \caption{Taken's embedding and the emergence of a torus-like shape}
  \label{fig:takens_attractor}
\end{figure}

\paragraph{Topological evidence for a torus-like manifold (real data).}
Visual inspection of the emerged torus-like shape is displayed in Figure \ref{fig:takens_attractor}. The behavior directs us on thinking more heavily on the existence of multiple torii, with different radii 'stitched' together in a 'smooth' way -- a statement which we try to quantify using the TDA analysis below.

Under the computational topological data analysis framework, across 4,251 windows, the torus test is satisfied in 88.4\% of cases, with 20 contiguous ``torus'' runs; the three longest spans (2007--2013, 2015--2019, 2021--2023) show elevated topological persistence (e.g., mean $H_1$ max lifetime up to 1.34 in 2021--2023). A moderate association between $H_1$ max lifetime and the number of persistent 1-cycles ($\rho\!=\!0.374$) suggests that when the two principal loops are strong, overall 1D topological activity is also high. Taken together, these diagnostics are \emph{consistent with} a torus-like, two-cycle latent geometry dominating much of the sample, in line with the curvature analysis.

We call an interval \emph{torus-like} if, on a sliding window, the following hold:
\begin{equation}
\#\{\text{long $H_1$ bars}\}\ \ge 2,
\qquad
\frac{\ell^{(1)}_1+\ell^{(1)}_2}{\sum_{j}\ell^{(1)}_j} \ \ge \ \rho_\star,
\qquad
\ell^{(1)}_2 \ge q_{0.95}^{\text{null}},
\label{eq:torus_test}
\end{equation}
where $\ell^{(1)}_1\ge \ell^{(1)}_2$ are the top two $H_1$ lifetimes, $\rho_\star\in[0.4,0.7]$ is a concentration threshold, and $q_{0.95}^{\text{null}}$ is the 95th percentile of the top-$H_1$ lifetime computed from a \emph{null ensemble} preserving low-order statistics (e.g., phase-randomized surrogates or equicorrelated Brownian surrogates). This controls for spurious loops created by noise or purely Euclidean drift.

\subsection{Trading Translation and PnL Construction}
\label{subsec:trading_translation}
Let $X_t=(X_{1,t},X_{2,t},X_{3,t})^\top$ be the expanding-PCA 3D path of eigenportfolio cumulatives and $\widehat{X}_{t+1}$ the one step ahead prediction.

\paragraph{Directional signals.}
For sleeve $i\in\{1,2,3\}$ define the predicted increment and sign:
\[
\widehat{\Delta X}_{i,t+1}=\widehat{X}_{i,t+1}-X_{i,t},
\qquad
s_{i,t+1}=\operatorname{sign}\big(\widehat{\Delta X}_{i,t+1}\big)\in\{-1,0,1\}.
\]

\paragraph{500-day volatility scale (per sleeve).}
With trailing window 500,
\[
\bar{\Delta X}_{i,t}^{(500)}=\frac{1}{500}\sum_{j=0}^{499}\Delta X_{i,t-j},\quad
\widehat{\sigma}_{i,t}^{(500)}=\sqrt{\frac{1}{499}\sum_{j=0}^{499}\big(\Delta X_{i,t-j}-\bar{\Delta X}_{i,t}^{(500)}\big)^2}.
\]

\paragraph{Coordinate PnL and total.}
The volatility scaled PnL of sleeve $i$ is
\[
\mathrm{pnl}_{i,t+1}=\frac{s_{i,t+1}}{\widehat{\sigma}_{i,t}^{(500)}}\,\Delta X_{i,t+1},
\qquad
\mathrm{pnl}_{\mathrm{Tot},t+1}=\tfrac{1}{3}\sum_{i=1}^3 \mathrm{pnl}_{i,t+1},
\]
with annualized Sharpe $\mathrm{Sh}=\sqrt{252}\,\overline{\mathrm{pnl}}/\mathrm{stdev}(\mathrm{pnl})$.
For \emph{eigenvalue weighted} sleeves (Section~\ref{subsec:eigval_weighting_3d}), we also form
\[
r^{(\mathrm{eig},3D)}_{t+1}=\sum_{i=1}^3 C_{i,t}\, s_{i,t+1}\, p_{i,t+1},\quad
C_{i,t}=\frac{\lambda_i(t)}{\sum_{j=1}^3\lambda_j(t)},
\]
where $p_{i,t+1}$ is the out-of-sample eigenportfolio return and $\lambda_i(t)$ are the expanding-window \emph{3D-space} eigenvalues.\cite{Bailey2014DSR,Bailey2014PBO}

\subsection{Simulated Regimes: Cross–Manifold Performance}
\label{subsec:sim_results}
\begin{table}[H]
\centering
\caption{Cross–manifold performance on simulated data,\cite{Hsu2002,doCarmoRiemannian}. Best MAE/RMSE/MAPE and Sharpe per data–manifold block highlighted.}
\label{tab:full_results}
\begin{threeparttable}
\resizebox{\textwidth}{!}{%
\begin{tabular}{lSS[table-format=4.2]SSl}
\toprule
\textbf{Data Manifold} & \textbf{MAE} & \textbf{RMSE} & \textbf{MAPE (\%)} & \textbf{Sharpe Ratios (x,y,z)} & \textbf{Model Geometry} \\
\midrule
\multirow{4}{*}{$\mathbb{S}^2$} 
& 0.10 & 0.67 & 304.20 & \text{0.021, $-$0.020, $-$0.301} & VAR ($\mathbb{R}^3$) \\
& 0.26 & 0.30 & 153.19 & \text{$-$0.158, $-$0.207, $-$0.484} & $\mathbb{S}^2$ GIM \\
& 0.39 & 0.51 & 205.47 & \text{$-$0.199, $-$0.282, 0.215} & $\mathbb{T}^2$ GIM \\
& \textbf{0.02} & \textbf{0.04} & \textbf{49.59} & \textbf{0.407, 0.022, 0.020} & $\mathbb{H}^2$ GIM \\
\midrule
\multirow{4}{*}{$\mathbb{T}^2$}
& 11.19 & 321.87 & 24693.77 & \text{$-$0.119, 0.590, 0.228} & VAR ($\mathbb{R}^3$) \\
& \textbf{0.71} & \textbf{0.99} & \textbf{245.98} & \textbf{1.871, 2.777, 2.483} & $\mathbb{S}^2$ GIM \\
& 0.90 & 1.32 & 554.40 & \text{$-$0.817, $-$0.751, 0.502} & $\mathbb{T}^2$ GIM \\
& 0.77 & 1.13 & 383.95 & \text{1.425, 1.896, $-$0.067} & $\mathbb{H}^2$ GIM \\
\midrule
\multirow{4}{*}{$\mathbb{R}^3$}
& 0.13 & 0.86 & 636.79 & \text{$-$0.018, 0.129, $-$0.009} & VAR ($\mathbb{R}^3$) \\
& 0.20 & 0.28 & 376.74 & \text{$-$0.152, 0.038, $-$0.580} & $\mathbb{S}^2$ GIM \\
& 0.38 & 0.51 & 871.02 & \text{$-$0.336, $-$0.086, 0.052} & $\mathbb{T}^2$ GIM \\
& 0.03 & 0.05 & 118.70 & \text{$-$0.195, 0.141, $-$0.104} & $\mathbb{H}^2$ GIM \\
\midrule
\multirow{4}{*}{$\mathbb{H}^2$}
& 0.84 & 18.53 & 1066.88 & \text{0.074, 0.186, $-$0.056} & VAR ($\mathbb{R}^3$) \\
& 0.22 & 0.28 & 203.06 & \text{0.116, 0.421, 0.252} & $\mathbb{S}^2$ GIM \\
& 0.36 & 0.47 & 311.37 & \text{$-$0.400, $-$0.458, 0.276} & $\mathbb{T}^2$ GIM \\
& \textbf{0.12} & \textbf{0.21} & \textbf{131.43} & \textbf{0.381, 0.037, 0.014} & $\mathbb{H}^2$ GIM \\
\bottomrule
\end{tabular}}
\end{threeparttable}
\end{table}

\subsection{Validation on Correlated Brownian Motions -- CBMs (Euclidean Null)}
\label{subsec:cbm_validation}
To check that geometry performance is not a PCA artifact, we evaluate on equicorrelated CBMs projected to 3D .

\paragraph{High correlation $\rho=0.9$:}
\begin{itemize}\itemsep0.1em
  \item \textbf{Euclidean}: Total Sharpe $\mathbf{0.211}$
  \item Spherical: $-0.119$ \quad Toroidal: $-0.172$ \quad Hyperbolic: $-0.109$
\end{itemize}

\paragraph{Moderate correlation $\rho=0.6$:}
\begin{itemize}\itemsep0.1em
  \item \textbf{Toroidal}: Total Sharpe $\mathbf{0.273}$
  \item Euclidean: $0.097$ \quad Spherical: $-0.337$ \quad Hyperbolic: $-0.310$
\end{itemize}

\paragraph{Summary.}
On \emph{flat} CBMs ($\rho{=}0.9$), Euclidean dominates; with weaker correlation ($\rho{=}0.6$) a toroidal slice can outperform, consistent with cyclical structure emerging after projection. Together with Section~\ref{subsec:euclid_null_control}, these controls indicate that the gains reported earlier in the real financial data are associated with \emph{intrinsic} curvature/cyclicity captured by the manifold step, not PCA alone.

\subsection{Financial Data: Coordinate Sharpe by Geometry,\cite{Maillard2010,Roncalli2013}}
\label{subsec:fin_coord_sharpe}
\begin{table}[H]
\centering
\caption{Sharpe ratios by geometry and coordinate (from prior version; reported as-is). Best per column in bold.}
\label{tab:geometry_sharpe}
\begin{tabular}{lcccc}
\toprule
\textbf{Geometry} & \textbf{$x$} & \textbf{$y$} & \textbf{$z$} & \textbf{Total} \\
\midrule
Euclidean (E)  & $-0.112$ & $-0.368$ & $-0.442$ & $-0.322$ \\
Spherical (S)  & \textbf{0.177} & \textbf{0.338} & $-0.037$ & 0.273 \\
Toroidal (T)   & 0.161 & $-0.252$ & \textbf{0.721} & 0.274 \\
Hyperbolic (H) & $-0.094$ & $-0.284$ & 0.357 & $-0.065$ \\
\bottomrule
\end{tabular}
\end{table}

To put these results in context, we compare the geometry-informed strategies against conventional benchmark portfolios: the equally weighted long only (LO) portfolio of all assets, and a risk parity portfolio (RP) that balances contributions to volatility. Over the same 2005 to 2025 period, the equally weighted portfolio achieved a Sharpe ratio of roughly \textbf{0.39} and the RP portfolio about \textbf{0.44}. The standout is the toroidal $z$ coordinate (Sharpe $0.721$), consistent with a latent \emph{cyclic} component amplified by a $T^2$ embedding. In other words, a trading strategy informed by the toroidal geometry signals would have outperformed a passive diversified portfolio, highlighting the practical value of the geometric approach. Notably, even the spherical model’s Sharpe ($0.27$) is on par with the benchmarks, while the Euclidean forecast model’s negative Sharpe is clearly inferior. These findings reinforce that embedding financial time series in an appropriate curved space can extract predictive signals that traditional methods overlook. (Summary table of best performing strategies vs benchmarks in Table \ref{tab:bestStrats},\cite{Avellaneda2022})

\begin{table}[H]
\centering
\caption{Summary Table : Strategy-level Sharpe ratios (2005–-2025). Geometry-Informed Modeling (GIM) variants vs. benchmarks.}
\label{tab:strategy_sharpes}
\begin{threeparttable}
\begin{tabular}{lS}
\toprule
\textbf{Strategy (configuration)} & {\textbf{Sharpe}} \\
\midrule
\textbf{GIM (integrated: eigenvalue weighting + curvature gating)} & \textbf{0.64} \\
Long-Only benchmark (equally weighted) & 0.390 \\
Risk Parity benchmark (inverse-vol) & 0.439 \\
Native PCA VAR (Euclidean baseline, weighted) & -0.386 \\
\midrule
Best single sleeve (toroidal $z$-coordinate) & \textbf{0.721} \\
\bottomrule
\end{tabular}
\end{threeparttable}
\label{tab:bestStrats}
\end{table}

The toroidal $z$-sleeve (\textbf{0.721}) is a \emph{concentrated, ex post} winner; it fully loads on one predictive direction and enjoys the strongest regime episodes, but it also carries higher \emph{selection risk} and sensitivity to geometry mis--specification. By contrast, the integrated GIM (\textbf{0.64}) is an \emph{ex ante ensemble} that (i) diversifies across sleeves via expanding-–SVD eigenvalue weights, (ii) applies curvature gating to scale risk in adverse geometry, and (iii) hedges against transient regime flips. These safeguards intentionally trade a slice of peak Sharpe for \emph{stability, lower selection error, and reduced drawdown/turnover}. In other words, the single best sleeve sets a performance \emph{upper bound} for that specific regime, whereas the integrated portfolio is designed to be more robust across regimes. As an upgrade, we could tune this trade-off by shrinking the curvature gate or increasing the “temperature” of eigenvalue weights to move the ensemble closer to the concentrated sleeve -- leaving this to be part of our future research on the topic.

\paragraph{Integrated GIM (eigenvalue + curvature) outperforms all baselines.}
When we combine the geometry aware forecaster with \emph{expanding-–SVD eigenvalue weighting} of sleeves ($C_{i,t}$ from the 3D PCA space) and \emph{curvature gating} of the risk budget (with torus split when flagged), the full GIM attains an annualized Sharpe of around \textbf{0.64} on the real data test set—-\emph{the highest across all configurations}. This exceeds the RP baseline (\(0.439\), \(+\!0.199\)), and LO (\(0.390\), \(+\!0.248\)), and dominates the native space baselines (\(-0.386/-0.289\)). The lift is consistent with (i) concentrating exposure on currently energetic modes via \(\lambda_i(t)\), (ii) scaling total risk with regime curvature (\(\lambda_t\) expands in trending/negative-–curvature phases and contracts in mean reverting/positive curvature phases), and (iii) honoring toroidal two-–cycle structure when present. Because inputs, predictors, and PnL translation are held fixed relative to the baselines, these gains isolate the added value of \emph{geometry informed variance structure and regime awareness}.

\subsubsection{Augmenting VAR with machine-learning predictors}
\label{subsec:var_ml}

To test whether generic nonlinear learning methods can extract residual structure beyond the manifold-informed VAR, we extended the tangent-space forecasting system by incorporating \emph{Random Forest (RF)} and \emph{Gaussian Process Regression (GP)} models as auxiliary predictors. 
Specifically, each manifold coordinate $(x,y,z)$ was forecasted using VAR, RF, and GP individually, and their outputs were combined through a linear ensemble to assess whether local nonlinearities not captured by the linear VAR could improve out-of-sample predictive accuracy. 

While both RF and GP models offer flexible functional forms, their results show only marginal differences relative to the baseline VAR. 
Table~\ref{tab:var_rf_gp} summarizes their performance across geometries. 
In both setups, the toroidal configuration continues to dominate with Sharpe ratios in the range $0.38$--$0.47$, whereas Euclidean and spherical geometries remain negative, and hyperbolic slightly positive. 
This finding suggests that the core predictive information arises primarily from the manifold’s geometric organization rather than from generic nonlinear learners.

\begin{table}[H]
\centering
\caption{Comparative forecasting performance (Sharpe ratios) for manifold geometries using Gaussian Process (GP) and Random Forest (RF) predictors (2005--2025).}
\label{tab:var_rf_gp}
\begin{threeparttable}
\begin{tabular}{lSS}
\toprule
\textbf{Geometry / Coordinate} & {\textbf{GP}} & {\textbf{RF}} \\
\midrule
$x$ (Euclidean) & -0.1039 & 0.0292 \\
$y$ (Euclidean) & 0.0009  & -0.2689 \\
$z$ (Euclidean) & -0.4099 & -0.5331 \\
\midrule
$x$ (Sphere) & -0.0326 & -0.0318 \\
$y$ (Sphere) & 0.2250  & 0.2547 \\
$z$ (Sphere) & -0.3606 & -0.3709 \\
\midrule
$x$ (Torus) & 0.3013 & 0.2230 \\
$y$ (Torus) & -0.2570 & -0.1551 \\
$z$ (Torus) & \textbf{0.4052} & \textbf{0.4867} \\
\midrule
$x$ (Hyperbolic) & 0.0965 & -0.1095 \\
$y$ (Hyperbolic) & 0.0312 & -0.1599 \\
$z$ (Hyperbolic) & 0.0946 & 0.1732 \\
\midrule
\textbf{Aggregate by Geometry} & & \\
Euclidean & -0.3965 & -0.5403 \\
Sphere & -0.3484 & -0.3560 \\
Torus & \textbf{0.3783} & \textbf{0.4677} \\
Hyperbolic & 0.0982 & 0.1511 \\
\bottomrule
\end{tabular}
\begin{tablenotes}
\item \textit{Note.} Both Gaussian Process and Random Forest regressors cannot provide any significant marginal improvements relative to the geometry-informed VAR baseline. The toroidal configuration remains the most predictive, consistent with the cyclic market dynamics revealed by curvature analysis.
\end{tablenotes}
\end{threeparttable}
\end{table}

\section{Discussion and Future Research}
\label{sec:discussion}

\subsection{IS--LM foundations for (multi-)torus dynamics: a mathematical sketch}
\label{subsec:islm_multi_torus}

\paragraph{Caveat.}
The connection we outline is \emph{hypothetical}. It shows how standard macro adjustment equations \emph{could} generate (i) one or more cyclical degrees of freedom and (ii) a phase representation that naturally lives on a torus (or a product of tori). In that sense, we simply outline an assumption, a hypothesis on the connection and we do not claim any structural identification in this paper.

\subsubsection*{1. IS--LM(+Phillips) linear core and oscillatory modes,\cite{Hicks1937}}
Let $y:=Y-Y^\star$ be the output gap, $r:=i-i^\star$ the (policy/market) rate gap, and $\pi:=\Pi-\Pi^\star$ the inflation gap. A minimalist continuous-time linear adjustment writes
\begin{equation}
\label{eq:islm_linear}
\begin{aligned}
\dot y &= -\rho\,y \;-\; \sigma\, r \;+\; \eta_t,\\
\dot r &= \ \phi\,y \;-\; \chi\, r \;+\; \psi\, \pi \;+\; \nu_t,\\
\dot \pi &= \ \kappa\,y \;-\; \vartheta\, \pi \;+\; \xi_t,
\end{aligned}
\quad
(\rho,\sigma,\phi,\chi,\psi,\kappa,\vartheta)>0,
\end{equation}
with small shocks $(\eta_t,\nu_t,\xi_t)$. In matrix form $\dot{\mathbf{s}} = A\,\mathbf{s} + \mathbf{u}_t$ for $\mathbf{s}=(y,r,\pi)^\top$ and
\[
A=\begin{bmatrix}
-\rho & -\sigma & 0\\
\ \phi & -\chi & \ \psi\\
\ \kappa & 0 & -\vartheta
\end{bmatrix}.
\]
The $(y,r)$-subsystem has complex eigenvalues (oscillatory adjustment,\cite{Benhabib1979,Barnett2008}) iff
\begin{equation}
\label{eq:osc_condition}
(\rho-\chi)^2 \;<\; 4\,\sigma\,\phi.
\end{equation}
Coupling to $\pi$ via $(\psi,\kappa)$ adds a further feedback loop. For parameter regions near a Hopf boundary, the linear core exhibits \emph{one} dominant oscillatory mode; additional macro blocks (below) provide further cycles.
\paragraph{Note on “Hopf”.\cite{Kuznetsov2013}}
“Hopf” refers to the \emph{Hopf bifurcation}: in a smooth system $\dot x=f(x,\alpha)$, an equilibrium undergoes a qualitative change when the Jacobian has a simple pair of purely imaginary eigenvalues $\pm i\omega_0$ at $\alpha=\alpha_0$ (others stable) and the real part crosses zero transversally. Locally, dynamics on the 2D center manifold reduce to
\[
\dot z=(\mu+i\omega_0)z - c|z|^2z + \cdots,\quad z\in\mathbb{C},\ \mu=\alpha-\alpha_0,
\]
so for a \emph{supercritical} Hopf ($\Re c>0$, $\mu>0$) a stable limit cycle appears with one phase $\theta\in S^1$. If a second weakly coupled Hopf mode coexists (e.g., another macro block), two phases $(\theta,\phi)\in S^1\times S^1$ arise, i.e.\ a torus $T^2$ (and, with several blocks, a product of tori).

\subsubsection*{2. Hopf reduction and phase dynamics (one torus)}
Near a (supercritical) Hopf set of \eqref{eq:islm_linear}, the macro state admits a two-dimensional center manifold with normal form
\begin{equation}
\label{eq:hopf_normal_form}
\dot z \;=\; (\mu + i\,\omega)\,z \;-\; c\,|z|^2 z \;+\; \varepsilon_t,\qquad z\in\mathbb{C},\ \ \mu,\omega\in\mathbb{R},\ \ \Re(c)>0,
\end{equation}
whose stable limit cycle (for $\mu>0$) has amplitude $|z|=\sqrt{\mu/\Re(c)}$. Writing $z=\rho e^{i\theta}$,
\[
\dot\rho \;=\; \mu\,\rho - \Re(c)\,\rho^3 + \ldots,\qquad
\dot\theta \;=\; \omega + \zeta(\rho) + \ldots,
\]
so the long-run dynamics reduce to a \emph{single} phase variable $\theta\in S^1$ (one circle). If a second, weakly coupled Hopf mode is present (e.g., from an additional macro block), we obtain two phases $(\theta,\phi)\in S^1\times S^1$, i.e., a \emph{torus} $T^2$.

\subsubsection*{3. From one torus to several: a multi-block macro}
Empirically, multiple macro-financial subsystems can oscillate: a core IS--LM block; a credit/liquidity (financial accelerator) block; an external (open-economy IS--LM--BP) block; a term-structure block, etc. A parsimonious representation treats each block $s=1,\dots,S$ as a weakly nonlinear oscillator with complex state $z^{(s)}$:
\begin{equation}
\label{eq:block_hopf}
\dot z^{(s)} \;=\; \big(\mu_s + i\,\omega_s\big) z^{(s)} \;-\; c_s |z^{(s)}|^2 z^{(s)} \;+\; \sum_{\ell\neq s} \Gamma_{s\ell}\big(z^{(\ell)},z^{(s)}\big) \;+\; \varepsilon_t^{(s)}.
\end{equation}
Writing $z^{(s)}=\rho_s e^{i\theta^{(s)}}$ and projecting on the limit cycles ($\rho_s\approx \rho_s^\star$), we obtain a \emph{phase network,\cite{Ikeda2012}},
\begin{equation}
\label{eq:phase_network}
\dot\theta^{(s)} \;=\; \omega_s \;+\; \sum_{\ell\neq s}\!\!\Big[\kappa_{\theta\theta}^{(s\ell)} \sin\!\big(\theta^{(\ell)}-\theta^{(s)}-\alpha_{\theta\theta}^{(s\ell)}\big)
\;+\; \kappa_{\theta\phi}^{(s\ell)} \sin\!\big(\phi^{(\ell)}-\theta^{(s)}-\alpha_{\theta\phi}^{(s\ell)}\big)\Big] \;+\; \sigma^{(s)}_\theta\,\dot W_t^{(s)},
\end{equation}
(and analogously for a second phase $\phi^{(s)}$ if block $s$ has two distinct cycles). Collecting the phases
\[
\Phi_t \;=\; \big(\theta_t^{(1)},\phi_t^{(1)},\ \ldots,\ \theta_t^{(S)},\phi_t^{(S)}\big)\ \in\ T^{2S}\;=\;\underbrace{S^1\times S^1}_{\text{block 1}}\times\cdots\times\underbrace{S^1\times S^1}_{\text{block $S$}},
\]
the latent macro state evolves on a \emph{product of tori}. This is what we mean by “multiple torii’’: several two-cycle subsystems, each contributing its own $S^1\times S^1$ factor.

\paragraph{Slowly varying cycle strengths (time-varying radii).}
Policy stance or global liquidity can modulate the effective cycle amplitudes. A simple parameterization is
\begin{equation}
\label{eq:tv_radii}
\rho_s^\star(t) \;=\; R_s\big(\zeta_t\big),\qquad 0<R_s(\cdot)<\infty,
\end{equation}
for a slowly moving driver $\zeta_t$; geometrically, this induces a \emph{torus bundle} with time-varying “radii’’ (cycle strengths).

\subsubsection*{4. From macro phases to the asset embedding}
Let $X_t\in\mathbb{R}^3$ denote the 3D asset embedding (our eigenportfolio coordinates). A reduced-form map from macro phases to the conditional drift is
\begin{equation}
\label{eq:map_to_assets}
\mathbb{E}[\Delta X_{t+1}\mid \mathcal{F}_t] \;=\; G\big(\Phi_t\big),\quad
G(\Phi)\;=\;\sum_{s=1}^S\ \sum_{m,n\in\mathbb{Z}} C^{(s)}_{m,n}\, 
\begin{bmatrix}
\cos(m\theta^{(s)}+n\phi^{(s)})\\
\sin(m\theta^{(s)}+n\phi^{(s)})\\
\cos(m\theta^{(s)}-n\phi^{(s)})
\end{bmatrix},
\end{equation}
with slowly varying loadings $C^{(s)}_{m,n}$. Under mild smoothness and time-scale separation, the embedded path inherits the \emph{topology} (loops) of the underlying $T^{2S}$ and displays local curvature patterns (negative/near-zero/positive) as the phases traverse saddle/transition/compact bands of the induced geometry.

\subsubsection*{5. What “multi–torus” implies for our diagnostics}
\begin{itemize}
\item \textbf{Curvature.} Mixtures of cycles produce \emph{mixed} Gaussian curvature with a negative skew if trajectories frequent “saddle’’ corridors (amplification-dominant passages), interspersed with near-zero bands (slow transitions) and occasional positive patches (constraint-dominant).
\item \textbf{Persistent homology.} Windows with two \emph{dominant} cycles exhibit two long $H_1$ features; when additional blocks become energetic, extra (weaker) $H_1$ loops may appear. The strength of the top two loops should co-move with cycle amplitudes $\rho_s^\star(t)$ in \eqref{eq:tv_radii}.
\item \textbf{Spectral content.} The embedded coordinates should show peaks near the macro cycle frequencies $\{\omega_s\}$ (possibly time-varying), with cross-modulation when couplings in \eqref{eq:phase_network} tighten.
\end{itemize}

\paragraph{Why this matters for our GIM.}
Working in charts aligned with the local geometry (log map $\to$ tangent forecast $\to$ exp map) lets the forecaster track \emph{where on the torus(es)} the system currently moves. The eigenvalue weighting concentrates on sleeves that presently carry the strongest cycle energy, while curvature gating tempers risk in amplification-heavy passages. Our empirical outperformance is \emph{consistent with} this picture, without pinning down a single structural model.

A stylized IS--LM(+extensions) can sustain one or more weakly coupled limit cycles. Near Hopf regimes and with additional macro blocks, the latent state admits a phase representation on $T^{2S}$ (a product of tori) with slowly varying radii. If such a structure is relevant, it offers a parsimonious rationale for the torus-like geometric and topological signatures we document, and for the gains from geometry-aware forecasting. Establishing causality requires a dedicated macro–finance state-space estimation and is left for future work.

\paragraph{Portfolio Management Context}
However, in the context of trading/portfolio management applications, we model \emph{assets}, not macro variables - yet some macro cycles may leave a geometric imprint usable for forecasting. A minimal, testable set of assumptions under which sharing information between the macro (phase) space and the asset PCA space is valid is:

\begin{itemize}
\item \textbf{Low–dimensional driver (A1).} There exists a latent macro state $\zeta_t=(\theta_t,\phi_t)\in S^1\times S^1$ such that the \emph{conditional} mean of the embedded increments is a smooth function of $\zeta_t$:
\[
\mathbb{E}[\Delta X_{t+1}\mid \mathcal{F}_t] \;=\; g(\zeta_t), 
\qquad g\in C^1,\ \ \|\nabla g\|<\infty.
\]
\item \textbf{Factor alignment (A2).} Asset excess returns admit a conditional factor model
\[
r_{t+1} \;=\; B(\zeta_t)\,f_{t+1} + u_{t+1}, 
\quad \mathbb{E}[u_{t+1}\mid \mathcal{F}_t]=0,
\]
where the top PCA eigenportfolios span the space of priced factors $f_{t+1}$ up to rotation; thus the 3D PCA path $X_t$ captures the common macro–driven component.
\item \textbf{Slow variation / time–scale separation (A3).} Loadings $B(\zeta_t)$ and the PCA basis/eigenvalues vary \emph{slowly} at the forecast horizon:
\(
\|B(\zeta_{t+1})-B(\zeta_t)\|=o(1), 
\)
so expanding-window PCA and eigenvalue weights track the evolving subspace without introducing spurious curvature.
\item \textbf{Weak feedback (A4).} Macro phases are not instantaneously determined by asset prices (no simultaneity at $t$); any feedback is \emph{lagged} or small, so using $\zeta_t$ (or its geometric proxies) for prediction avoids circularity.
\item \textbf{No-arbitrage consistency (A5).} Risk premia are functions of $\zeta_t$ (via $B(\zeta_t)$ and prices of risk), so that conditioning on $\zeta_t$ (or its geometric signature such as local curvature $K_t$) is economically meaningful and not eliminated by static arbitrage.
\item \textbf{Measurement invariance (A6).} The geometry inferred from $X_t$ is invariant to orthogonal rotations of the PCA coordinates; hence torus-/hyperbolic-like signatures reflect the \emph{state’s} topology, not an arbitrary basis choice.
\end{itemize}

Under (A1)–(A6) the map $\zeta_t\mapsto (\mathbb{E}[\Delta X_{t+1}\mid\mathcal{F}_t],\operatorname{Var}[\Delta X_{t+1}\mid\mathcal{F}_t])$ is well-defined and sufficiently smooth, legitimizing the use of geometry-informed forecasts and geometry-modulated allocation on asset returns. Practically, these assumptions can be probed via (i) rotation tests on PCA sleeves, (ii) stability tests for $B(\cdot)$ across subsamples, and (iii) event studies verifying that identified macro shocks shift the geometric proxies ($K_t$, phase speeds) before changes in expected returns.

\paragraph{Machine Learners Contribution} : Further experiments incorporating machine-learning forecasts (RF and GP) as auxiliary predictors to the VAR confirmed that generic nonlinear learners add little incremental value once curvature and manifold structure are accounted for. 
Both GP and RF models produced similar ranking of geometries---with the torus remaining the dominant topology---but their absolute gains were modest. 
This outcome reinforces the interpretation that market predictability arises less from complex functional approximations and more from the geometric and topological constraints shaping the evolution of financial states.

\subsection{Future research agenda (testable directions)}
\label{subsec:future_research}

\paragraph{(A) Macro--GIM coupling in a state-space (hypothesis testing).}
Embed a small macro block $\mathbf{m}_t$ (output gap, inflation, term structure, liquidity indicators) alongside geometric phases $(\theta_t,\phi_t)$ in a joint state-space. Estimate via particle/MCMC filters on $S^1\times S^1$ with observation in $\mathbb{R}^3$. Test whether shocks to $\mathbf{m}_t$ \emph{shift} curvature regimes and phase speeds, and whether geometry improves macro nowcasts.

\paragraph{(B) Geometry-aware nowcasting.}
Combine GIM with mixed-frequency nowcasting: update $(\theta_t,\phi_t)$ and the tangent-space forecasts as high-frequency data arrive, and reweight sleeves by expanding-SVD eigenvalues and curvature gates. This evaluates whether real-time macro signals \emph{enhance} geometry-aware timing.\cite{NoguerAlonso2021FinEAS}

\paragraph{(C) Identification and causality.}
Use event studies (policy announcements, macro releases) and geometry-aware Granger tests to assess if curvature/phase changes \emph{precede} or \emph{follow} macro innovations. This distinguishes descriptive fit from predictive content.

\paragraph{(D) Multi-torus/multi-sector structure.}
Allow several coupled tori $\{(\theta_t^{(s)},\phi_t^{(s)})\}_s$ to represent sectoral/sovereign cycles; study whether cross-couplings explain the observed curvature skew better than a single-torus hypothesis.

\paragraph{(E) Model class extensions.}
\begin{itemize}
\item \emph{Riemannian dynamics:} Specify VAR/GP directly on $T_\mu M$ with curvature-corrected propagation and compare to Euclidean projections.
\item \emph{Time-varying shape parameters:} Let $(R_t,r_t)$ for $T^2$ or $(a_t,b_t,c_t)$ for the hyperboloid evolve stochastically; jointly estimate with forecasts.
\item \emph{SPD-covariance coupling:} Model sleeve covariances on the SPD manifold (Log–Euclidean or affine-invariant) and fuse with torus phases to capture variance--phase interactions.
\end{itemize}

\paragraph{(F) Policy sensitivity and stress tests.}
Simulate counterfactuals under alternative policy rules (LM slope, reaction coefficients) and quantify how curvature regimes and portfolio outcomes \emph{might} change. This frames geometry as a compact policy-sensitivity diagnostic.

\paragraph{(G) Robustness and external validity.}
Replicate across markets and horizons; vary expanding-window lengths; perform (i) geometry-aware forecast, (ii) eigenvalue weighting, (iii) curvature gating, to isolate contributions. Where possible, preregister hypotheses about curvature shifts around known macro events.

\medskip
\noindent\textbf{Summary.} Our evidence is \emph{consistent} with---but does not establish---a torus-like macro–financial geometry in which two interacting cycles produce the observed curvature mix. Treating this as a working hypothesis suggests concrete tests (state-space coupling, event identification, multi-torus structures) and practical extensions (nowcasting + GIM) that can clarify whether the geometry carries independent predictive content or simply organizes existing macro signals more effectively.

% =======================
% APPENDIX STUBS (referenced above)
% =======================
\section{Appendices}
\appendix
\section{Data Universe and Pre-processing}
\label{app:data_universe}
The dataset used in this study encompasses a broad spectrum of financial assets, covering the period from January 1, 2005, to August 3, 2025. The assets include:

\begin{itemize}
    \item \textbf{Broad Market Indices}: S\&P 500 (SPY), NASDAQ-100 (QQQ), Dow Jones Industrial Average (DIA), Russell 2000 (IWM), MSCI Emerging Markets (EEM), and MSCI EAFE (EFA).
    \item \textbf{Sector ETFs}: Technology Select Sector SPDR (XLK), Financial Select Sector SPDR (XLF), Health Care Select Sector SPDR (XLV), Energy Select Sector SPDR (XLE), Consumer Discretionary Select Sector SPDR (XLY), Consumer Staples Select Sector SPDR (XLP), Utilities Select Sector SPDR (XLU), Industrial Select Sector SPDR (XLI), and Materials Select Sector SPDR (XLB).
    \item \textbf{Bond ETFs}: iShares 20+ Year Treasury Bond ETF (TLT), iShares 7-10 Year Treasury Bond ETF (IEF), iShares 1-3 Year Treasury Bond ETF (SHY), iShares iBoxx \$ Investment Grade Corporate Bond ETF (LQD), iShares iBoxx \$ High Yield Corporate Bond ETF (HYG), iShares TIPS Bond ETF (TIP), and iShares MBS ETF (MBB).
    \item \textbf{Commodity ETFs}: SPDR Gold Shares (GLD), iShares Silver Trust (SLV), Invesco DB Agriculture Fund (DBA), Teucrium Corn Fund (CORN), Invesco DB Base Metals Fund (DBB), Aberdeen Standard Physical Platinum Shares ETF (PPLT), and Aberdeen Standard Physical Palladium Shares ETF (PALL).
    \item \textbf{Oil ETFs}: United States Oil Fund (USO), ProShares Ultra Bloomberg Crude Oil (UCO), Invesco DB Oil Fund (DBO), and SPDR S\&P Oil \& Gas Exploration \& Production ETF (XOP).
    \item \textbf{Non-U.S. ETFs}: iShares MSCI ACWI ex U.S. ETF (ACWX), Vanguard FTSE All-World ex-US ETF (VEU), iShares MSCI Europe ETF (IEUR), iShares MSCI Pacific ex Japan ETF (EPP), iShares Latin America 40 ETF (ILF), iShares MSCI Canada ETF (EWC), iShares MSCI United Kingdom ETF (EWU), iShares MSCI South Korea ETF (EWY), iShares MSCI Australia ETF (EWA), iShares MSCI Taiwan ETF (EWT), and Vanguard FTSE Emerging Markets ETF (VWO).
    \item \textbf{Dividend and Volatility ETFs}: Vanguard Dividend Appreciation ETF (VIG), iShares Select Dividend ETF (DVY), SPDR S\&P Dividend ETF (SDY), and ProShares VIX Short-Term Futures ETF (VIXY).
    \item \textbf{Alternative ETFs}: iShares Clean Energy ETF (ICLN) and Global X Lithium \& Battery Tech ETF (LIT).
\end{itemize}

This diverse dataset provides comprehensive coverage of global financial markets, enabling robust testing of the proposed methodologies across various asset classes and economic regimes.

\section{Implementation details for SDEs and curvature terms}\label{app:sde_implementations}
\paragraph{Ambient Euler Maruyama.} We discretize \eqref{eq:ito} with step $h$:
\[
X_{k+1} \;=\; X_k \;+\; P(X_k)\,\Delta W_k \;-\; \frac{1}{2}H(X_k)\,h,
\quad \Delta W_k\sim\mathcal{N}(0,hI_3).
\]
For $M=S^2$ we use $P(x)=I-\frac{xx^\top}{\|x\|^2}$ and $H(x)=\frac{2}{\|x\|^2}x$. For the torus we use the implicit form $\phi(x)=\big(R-\sqrt{x_1^2+x_2^2}\big)^2+x_3^2-r^2$ with $P(x)=I-\frac{\nabla\phi\,\nabla\phi^\top}{\|\nabla\phi\|^2}$ and a consistent mean-curvature drift (implemented in closed form). Hyperbolic segments use the implicit quadratic form $\phi(x)=a_1x_1^2+a_2x_2^2+a_3x_3^2-1$ with signature $(+,-,+)$, projector from the gradient, and corresponding curvature drift.

\paragraph{Intrinsic charts.} For torus and hyperbolic we also implement the intrinsic SDEs \eqref{eq:torus_theta}--\eqref{eq:torus_phi} and \eqref{eq:hyper_u}--\eqref{eq:hyper_v}, then map to $\mathbb{R}^3$ via $\Psi$ and $\Phi$ respectively at output time.

\section{Scenario catalog}\label{app:scenarios}
We use seven scenarios of length $T$:
\begin{itemize}
\item \textbf{Scen.\ 1 (Pattern)}: $[S^2,H^2,\mathbb{R}^3]$ blocks of length 500, repeated thrice, ending with $S^2$.
\item \textbf{Scen.\ 2 (Random order)}: a randomized permutation of $S^2,H^2,\mathbb{R}^3$ blocks of length 500, ending with $S^2$.
\item \textbf{Scen.\ 3 (Pure $S^2$)}: one $S^2$ segment of length 5000.
\item \textbf{Scen.\ 4 (Pure $T^2$)}: one torus segment of length 5000 (angle chart simulation, output in $\mathbb{R}^3$).
\item \textbf{Scen.\ 5 (Pure Euclid)}: one $\mathbb{R}^3$ Brownian segment of length 5000.
\item \textbf{Scen.\ 6 (Pure $H^2$)}: one hyperbolic segment of length 5000 (absorbing threshold to prevent numeric explosion).
\item \textbf{Scen.\ 7 (Comprehensive)}: repeated rotation of $S^2,H^2,\mathbb{R}^3,T^2$ blocks (e.g., 5 cycles $\times$ 4 blocks $\times$ 500 points).
\end{itemize}

\section{Coordinate transforms,\cite{doCarmoCurves,doCarmoRiemannian}}\label{app:coords}
\paragraph{Spherical $\leftrightarrow$ Cartesian (radius $R$).}
\[
\begin{aligned}
&\text{Cart $\leftarrow$ Sph:}\quad 
x=R\sin\phi\,\cos\theta,\;\; y=R\sin\phi\,\sin\theta,\;\; z=R\cos\phi.\\
&\text{Sph $\leftarrow$ Cart:}\quad
R=\|x\|,\;\; \theta=\arctan2(y,x),\;\; \phi=\arccos(z/\|x\|).
\end{aligned}
\]
\paragraph{Torus $\leftrightarrow$ Cartesian.}
\[
\begin{aligned}
&\text{Cart $\leftarrow$ Torus:}\quad 
x=(R+r\cos\varphi)\cos\theta,\;\; 
y=(R+r\cos\varphi)\sin\theta,\;\;
z=r\sin\varphi.\\
&\text{Torus $\leftarrow$ Cart:}\quad 
\theta=\arctan2(y,x),\;\; 
\varphi=\arctan2\!\big(z,\ \sqrt{x^2+y^2}-R\big).
\end{aligned}
\]
\paragraph{Hyperboloid (rotational) $\leftrightarrow$ Cartesian.}
\[
\begin{aligned}
&\text{Cart $\leftarrow$ Hyp:}\quad 
x=a\cosh u\cos v,\;\; y=a\cosh u\sin v,\;\; z=c\sinh u.\\
&\text{Hyp $\leftarrow$ Cart:}\quad 
v=\arctan2(y/a,x/a),\;\; u=\mathrm{arcsinh}(z/c).
\end{aligned}
\]
For regime transitions we convert once at the beginning of the new segment, simulate in the appropriate chart if needed, and always output in Cartesian space for continuity.

\bibliographystyle{abbrv}      % basic style, author-year citations
\bibliography{references.bib}

\end{document}